\documentclass[fleqn,usenatbib]{mnras}

\usepackage[T1]{fontenc}
\usepackage{ae,aecompl}

\usepackage{myaasmacros}
\usepackage{graphicx}	% Including figure files
\usepackage{amssymb}	% Extra maths symbols
\usepackage{booktabs,tabularx,graphicx,amsmath,makecell,hhline,enumitem,gensymb,multirow,ulem,pifont,adjustbox} %wasysym, fancyvrb
\usepackage[dvipsnames]{xcolor}
\usepackage{newtxtext,newtxmath}
\definecolor{Halo1}{HTML}{305FEA}
\definecolor{Halo2}{HTML}{00A982}
\definecolor{GM}{HTML}{EAA209}
\definecolor{GM2}{HTML}{000000}
\definecolor{Fcore}{HTML}{F800FF}
\definecolor{MO}{HTML}{228B22}

\title[The puzzling ellipticity of Eridanus II's star cluster]{EDGE: the puzzling ellipticity of Eridanus II's star cluster and its implications for dark matter at the heart of an ultra-faint dwarf}

\author{M. D. A. Orkney; J. I. Read; O. Agertz; A. Pontzen; M. Rey; M. Delorme]{Matthew D. A. Orkney$^1$, Justin I. Read$^{1}$\thanks{E-mail: justin.inglis.read@gmail.com}, Oscar Agertz, Andrew Pontzen, \\Martin P. Rey, Alex Goater, Ethan Taylor, Stacy Y. Kim Maxime Delorme}\\
$^1${\small Department of Physics, University of Surrey, Guildford, GU2 7XH, Surrey, UK}\\
}

\author[Matthew D. A. Orkney et al.] 
{Matthew D. A. Orkney$^{1,2}$,
Justin I. Read$^{1}$, 
Oscar Agertz$^{3}$, 
Andrew Pontzen$^{4}$, 
Martin P. Rey$^{5,3}$,
Alex Goater$^{1}$,
\newauthor
Ethan Taylor$^{1}$,
Stacy Y. Kim$^{1}$ and Maxime Delorme$^{6}$
\\
$^1$Department of Physics, University of Surrey, Guildford, GU2 7XH, United Kingdom\\
$^2$Institut de Ciencies del Cosmos (ICCUB), Universitat de Barcelona (IEEC-UB), Martí i Franquès 1, E08028 Barcelona, Spain\\
$^3$Lund Observatory, Department of Astronomy and Theoretical Physics, Lund University, Box 43, SE-221 00 Lund, Sweden\\
$^4$Department of Physics and Astronomy, University College London, London WC1E 6BT, UK\\
$^5$Sub-department of Astrophysics, University of Oxford, DWB, Keble Road, Oxford OX1 3RH, UK\\
$^6$Département d'Astrophysique/AIM, CEA/IRFU, CNRS/INSU, Université Paris-Saclay, 91191 Gif-Sur-Yvette, France}
\date{Accepted XXX. Received YYY; in original form ZZZ}

\pubyear{2020}

\begin{document}
\label{firstpage}
\pagerange{\pageref{firstpage}--\pageref{lastpage}}
\maketitle

\begin{abstract}
The Eridanus II (EriII) `ultra-faint' dwarf has a large ($15\,$pc) and low mass ($4.3\times10^3$\,M$_\odot$) star cluster (SC) offset from its centre by $23\pm3$\,pc in projection. Its size and offset are naturally explained if EriII has a central dark matter core, but such a core may be challenging to explain in a $\Lambda$CDM cosmology. In this paper, we revisit the survival and evolution of EriII's SC, focussing for the first time on its puzzlingly large ellipticity ($0.31^{+0.05}_{-0.06}$). We perform a suite of 960 direct $N$-body simulations of SCs, orbiting within a range of spherical background potentials fit to ultra-faint dwarf (UFD) galaxy simulations. We find only two scenarios that come close to explaining EriII's SC. In the first, EriII has a low density dark matter core (of size $\sim70\,\text{pc}$ and density $\lesssim2\times10^8\,\text{M}_{\odot}\,\text{kpc}^{-3}$). In this model, the high ellipticity of EriII's SC is set at birth, with the lack of tidal forces in the core allowing its ellipticity to remain frozen in for long times. In the second, EriII's SC orbits in a partial core, with its high ellipticity owing to its imminent tidal destruction. However, this latter model struggles to reproduce the large size of EriII’s SC, and it predicts substantial tidal tails around EriII’s SC that should have already been seen in the data. This leads us to favour the cored model. We discuss potential caveats to these findings, and the implications of the cored model for galaxy formation and the nature of dark matter. 
\end{abstract}


%%%%%%%%%%%%%%%%%%%%%%%%%%%%%%%%%%%%%%%%%%%%%%%%%%

%%%%%%%%%%%%%%%%% BODY OF PAPER %%%%%%%%%%%%%%%%%%

\section{Introduction}\label{sec:intro}

The Standard Cosmological Model ($\Lambda$CDM) provides an excellent match to the observed growth of cosmic structure on large scales \citep[e.g.][]{2006Natur.440.1137S, 2006ApJ...648L.109C, 2006PhRvD..74l3507T, 2013AJ....145...10D, 2014MNRAS.439.2515O, 2014A&A...571A..16P, 2016MNRAS.456.2301W}. However, pure dark matter structure formation simulations in a $\Lambda$CDM cosmology deviate from observations on galactic and sub-galactic scales, where `baryonic physics' -- e.g. gas cooling, star formation, `feedback' from massive stars -- becomes important \citep{1988ApJ...332L..33C, 1994ApJ...427L...1F, 1994Natur.370..629M, 1999ApJ...522...82K, 1999ApJ...524L..19M, 2001AJ....122.2381M, 2002ApJ...574L.129S, 2009MNRAS.399L.174O, 2011MNRAS.415L..40B}.
The earliest such tension has become known as the `cusp-core problem', and concerns the central densities of dark matter haloes in dwarf galaxies. Pure dark matter simulations in a $\Lambda$CDM cosmology predict the formation of a complex web of cosmic structure punctuated by dense haloes that have a self-similar density profile\footnote{This self-similarity is predicted to end at around an Earth-mass for a $\sim 100$\,GeV/c$^2$ Weakly Interacting Massive Particle due to the free-streaming limit \citep{2001PhRvD..64h3507H,2005Natur.433..389D,2020Natur.585...39W}. This is currently far below the resolution limit of most cosmological simulations.} \citep{navarro1997universal}. These density profiles are well-fit within an accuracy of ${\sim}10$ percent by the `NFW' profile \citep{1996ApJ...462..563N}, which is described by a divergent inner density that goes as $\rho \propto r^{-1}$ and outer density that goes as $\rho \propto r^{-3}$. \par

Analysis of real galaxies reveals a more complicated picture. Many dwarf galaxies favour a flatter central density than predicted by the NFW profile \citep{1988ApJ...332L..33C, 1994ApJ...427L...1F, 1994Natur.370..629M, 2001AJ....122.2381M, 2017MNRAS.467.2019R}, while others are consistent with a dense NFW cusp \citep{2014MNRAS.441.1584R, 2017ApJ...838..123S, 2018MNRAS.481..860R, 2021arXiv210101282S}. Recently, \citet{2019MNRAS.484.1401R} found that the densest dwarfs appear to be those with the least star formation, while the dwarfs with lower central densities have experienced more star formation. This is exactly what was predicted by models in which dark matter cusps are transformed to lower density cores by repeated stellar-feedback-induced gravitational potential fluctuations  \citep[e.g.][]{1996MNRAS.283L..72N, 2005MNRAS.356..107R,
2006Natur.442..539M,  2008Sci...319..174M, 2012MNRAS.421.3464P,2014MNRAS.437..415D}. Further dynamical evidence for dark matter cusp-core transformations in galaxies at higher redshift has been recently reported by \citet{genzel20}, \citet{bouche21} and \citet{sharma21}. \par

While gas cooling and stellar feedback can transform dark matter cusps to cores, it is energetically challenging for this process to create large dark matter cores (typically $>500$\,pc) in the very smallest  `ultra-faint' dwarfs, since they form so few stars ($M_* < 10^5$\,M$_\odot$; \citealt{2012ApJ...759L..42P, 2013MNRAS.433.3539G,2014MNRAS.437..415D,2015ApJ...806..229M,2015MNRAS.454.2092O,2016MNRAS.456.3542T}). However, smaller dark matter cores may still form inside the half-light radius ($R_{1/2} \sim 20-200$\,pc for UFDs\footnote{Whilst such cores would be small, they are nonetheless dynamically important by construction since $R_{1/2}$ is the scale on which we can probe the inner gravitational potential of dwarfs via their stellar kinematics \citep{2016MNRAS.459.2573R}.}) where the gravitational potential fluctuations are strongest \citep[e.g.][]{2015MNRAS.454.2092O,2016MNRAS.459.2573R}. Whether this is expected to happen in a $\Lambda$CDM cosmology remains an active area of debate. Most studies to date find that cusp-core transformations are challenging at the likely mass-scale of UFDs ($M_{200c} \sim 10^9$\,M$_\odot$; $M_* \sim 10^5$\,M$_\odot$; e.g. \citealt{2015MNRAS.454.2981C,wheeler19,gutcke21}). However, there are some notable exceptions. In recent work, \citet{orkney} studied cusp-core transformations for UFDs drawn from the `Engineering Dwarfs at Galaxy Formation's Edge' (EDGE) simulation project (with a mass, baryonic mass and spatial resolution of $120$\,M$_\odot$, $20$\,M$_\odot$ and $3$\,pc, respectively, sufficient to resolve dark matter cores in UFDs larger than ${\sim}20$\,pc). They found that UFDs can lower their inner dark matter density by up to a factor ${\sim}2$ through a combination of early heating due to star formation, followed by late-time heating from minor mergers. While none of their simulated dwarfs formed a completely flat central core, their small sample size left open the possibility that this combination of mechanisms could form flatter cores for some, rarer, assembly histories. It is also important to note that dark matter core formation on these mass scales is very sensitive to small changes in the star formation and feedback modelling. \citet{2020arXiv200903313P} showed that a small increase in variability of the star formation rate on time-scales shorter than the local dynamical time is sufficient to form a full dark matter core in an ultra-faint, without significantly altering its stellar mass. As such, the question of whether complete dark matter core formation is expected in some or all UFDs in $\Lambda$CDM remains open. \par

On the observational side, there is growing, albeit indirect, evidence for small dark matter cores in at least some UFDs \citep[e.g.][]{2018MNRAS.478.3879S,2017ApJ...844...64A,2020arXiv200512919M}, with the most compelling case being EriII \citep{2017ApJ...844...64A, 2018MNRAS.476.3124C}. Discovered by the Dark Energy Survey \citep[DES,][]{2015ApJ...807...50B, 2015ApJ...805..130K}, EriII is an UFD with a stellar mass of $8.3\substack{+1.7 \\ -1.4}\times10^{4}\,\text{M}_{\odot}$ and a dynamical mass estimate of $1.2\substack{+0.4 \\ -0.3}\times10^{7}\,\text{M}_{\odot}$ within its half-light radius \citep{2017ApJ...838....8L}. Its low metallicity implies an old stellar population and it is a good candidate for being a `fossil' galaxy that was quenched by reionisation \citep{2020arXiv201200043S}. Deep imaging of EriII has revealed a lone SC offset in projection from EriII's photometric centre \citep{2015ApJ...805..130K, 2016ApJ...824L..14C, 2020arXiv201200043S}, making it the faintest known galaxy to host a SC. The large projected half-light radius of this SC ($15\pm1$\,pc), and its large offset from EriII's photometric centre ($23\pm3$\,pc; \citealt{2020arXiv201200043S}\footnote{See also \citet{2021arXiv210901177M} who find an even larger offset and half-light radius, and a larger ellipticity.}), can be naturally explained if it is orbiting within a central dark matter core of size $\sim70\,\text{pc}$ and density $\lesssim2\times10^8\,\text{M}_{\odot}\,\text{kpc}^{-3}$  \citep{2017ApJ...844...64A,2018MNRAS.476.3124C}. By contrast, a `pristine cuspy' inner dark matter profile -- as expected in pure dark matter structure formation simulations in $\Lambda$CDM -- rapidly destroys the SC, making such a scenario unlikely. \par

More recently, kinematic analysis of spectroscopic observations from the MUSE-faint survey has confirmed that the SC in EriII is indeed a self-bound object and is within the host galaxy \citep{2020A&A...635A.107Z}. Expanding on this work, \citet{2021arXiv210100253Z} use mass-modelling techniques on 92 tracer stars to show that a solitonic core dark matter model provides the best match for the potential profile of EriII (but that alternative DM models cannot be ruled out). This, then, can be considered direct and independent evidence favouring a core in EriII, which serendipitously supports a core radius in excess of the projected orbital offset observed for the SC ($>47\,$pc at 68 percent confidence level). \par

In this paper, we revisit the survival and evolution of EriII's SCs, focussing for the first time on its puzzlingly large ellipticity ($0.31\substack{+0.05 \\ -0.06}$; \citealt{2020arXiv201200043S}). The ellipticity is a key new piece of information because, at first sight, it appears to pose an immediate problem for models in which the SC orbits within a constant density core. This is because: (i) the SC's short relaxation time ($t_{\rm relax} \sim 2$\,Gyrs; \citealt{2018MNRAS.476.3124C}) should rapidly erase any birth ellipticity through two-body effects \citep[e.g.][]{1958SvA.....2...22A,1976ApJ...210..757S, 1996A&A...308..453L,einsel99,bianchini13}; and (ii) it is challenging to induce a high ellipticity with tides, since tidal forces formally vanish inside a constant density core\footnotemark\ \citep[e.g.][]{2018MNRAS.476.3124C}. As such, we set out in this paper to answer the question: {\it can we form a SC as large, offset and elliptical as that in EriII within realistic UFDs in $\Lambda$CDM?} To address this question, we first select a number of `ultra-faint' dwarfs from the EDGE simulation project \citep{2019ApJ...886L...3R, 2020MNRAS.491.1656A, 2020arXiv200903313P, 2020MNRAS.497.1508R, orkney, 2021arXiv210700663P, 2021arXiv211203280R}, chosen to be most similar to EriII. We then set up a suite of 960 collisional {\sc nbody6df} follow-up simulations that model SCs orbiting within spherically symmetric potentials fit to these dwarfs, sampling over a grid of initial properties and orbits. We consider a static pristine cold dark matter cusp, a static weakened cusp, a static partial dark matter core, and a mildly time-evolving partial core, each fit to a representative UFD drawn from the EDGE simulation suite. We also include, for comparison, the lower density cored potential from \citet{2018MNRAS.476.3124C}. This is more cored than any of the EDGE dwarfs presented in \citet{orkney}, but has properties similar to the cored UFD presented and discussed in \citet{2020arXiv200903313P}. We discuss whether such a cored model is likely in a $\Lambda$CDM cosmology in \S\ref{discussion}. \par

\footnotetext{The tidal radius of a point mass SC of mass $M_{\rm SC}$ moving on a radial orbit in a power law background potential, $\rho \propto r^{-\gamma}$, is given by \citep[e.g.][]{Read06}: $\frac{r_{\rm t}}{r_{\rm B}} \simeq \left[\frac{M_{\rm SC}}{M_{\rm B}(1-\gamma)}\right]^{1/3}$, where $M_{\rm B}$ is the background mass enclosed within $r_{\rm B}$ (valid for $\gamma < 1$). Thus, for a constant density core, $\gamma = 0$, $r_{\rm t} \simeq r_{\rm B}$, and inside the core ($r < r_{\rm B}$) there will formally be no tidal stripping.}

This paper is organised as follows.
In \S\ref{method}, we describe the EDGE simulations used in this paper, and the motivation and setup for our collisional $N$-body simulation suite.
We discuss the results of our cosmological simulations in \S\ref{cosmo}, and present the results for our $N$-body simulation suite in \S\ref{nbody}.
Our discussion is presented in \S\ref{discussion}. In \S\ref{largecore}, we examine the core formation mechanisms in our EDGE simulations and consider whether alternative star formation and feedback physics, or alternative cosmologies could provide larger cores.
In \S\ref{pathways2} and \S\ref{pathways3}, we discuss possible pathways for the formation, evolution and survival of the SC in EriII.
Finally, in \S\ref{conclusions} we present our conclusions. \par

\section{Method} \label{method}

\begin{table*}
\resizebox{\textwidth}{!}{
\begin{tabular}{lcccccccc} 
\toprule
\textbf{Name} & \textbf{$M_{\rm 200c}$ [M$_{\odot}$]} & \textbf{$M_*$ [M$_{\odot}$]} & \textbf{$R_{1/2}$ [kpc]} & \textbf{[Fe/H]} & \textbf{$\sigma_{\rm LOS}$ [km\,s$^{-1}$]} & $\rho(40\,\text{pc})$ [M$_{\odot}/\text{kpc}^{3}$] & $M(<R_{1/2})$ [M$_{\odot}$] & \textbf{$t_{\rm form}$ [Gyr]} \\
\midrule
\textcolor{Halo1}{Halo1445} & $1.32\times10^9$ & $1.35\times10^5$ & 0.10 & -2.41 & $5.11\pm0.51$ & $4.78\times10^8$ & $9.93\times10^5$ & 3.79 \\
\textcolor{Halo2}{Halo1459} & $1.43\times10^9$ & $3.77\times10^5$ & 0.10 & -1.96 & $5.81\pm0.53$ & $6.89\times10^8$ & $1.91\times10^6$ & 2.83 \\
\textcolor{GM}{Halo1459 GM:Later} & $1.43\times10^9$ & $1.11\times10^5$ & 0.20 & -2.47 & $5.88\pm0.57$ & $3.43\times10^8$ & $2.92\times10^6$ & 3.19 \\
\textcolor{GM2}{Halo1459 GM:Latest} & $1.38\times10^9$ & $8.65\times10^4$ & 0.30 & -2.80 & $6.12\pm0.48$ & $3.62\times10^8$ & $7.26\times10^6$ & 3.55 \\
Eridanus II & 5.4$^{+5.6}_{-2.6}\times10^{8\:\mathrm{a}}$ & 8.3$^{+1.7}_{-1.4}\times10^{4\:\mathrm{b}}$ & 0.277$\pm0.014^{\:\mathrm{c}}$ & -2.38$\pm0.13^{\:\mathrm{d}}$ & 6.9$^{+1.2\:\mathrm{d}}_{-0.9}$ & $5.5^{+2.0}_{-2.2}\times10^{8\:\mathrm{a}}$ & $7.9^{+1.6}_{-1.5}\times10^{6\:\mathrm{a}}$ & - \\
\bottomrule
\end{tabular}
}
\caption{Comparison between the properties of EriII and those of our simulated galaxies at $z=0$. From left to right are the galaxy name, the virial mass, the total stellar mass, the projected half-light radius, the stellar metallicity within that radius, the stellar velocity dispersion, the galactic density at 40\,pc, the mass within the half-light radius using the mass estimator from \citet{2010MNRAS.406.1220W}, and the formation time, measured as the time when the main progenitor reaches 50 percent of its final $M_{\rm 200c}$. For the simulated dwarfs, the velocity dispersion and uncertainty are determined using a bootstrapping technique with $\sigma_{\rm \star}=\sqrt{\sigma^2_{\rm \star,x}+\sigma^2_{\rm \star,y}+\sigma^2_{\rm \star,z}}/\sqrt{3}$. Properties for EriII are gathered from the literature: a) \citealt{2021arXiv211209374Z} (their `core+tides' model); b) \citealt{2015ApJ...807...50B}; c) \citealt{2016ApJ...824L..14C}; d) \citealt{2017ApJ...838....8L}.}
\label{tab:eri_II_comparison}
\end{table*}

\subsection{Cosmological simulations of ultra-faint dwarfs}

We analyse a selection of UFD galaxy simulations taken from the wider suite of cosmological zoom-in simulations that form the EDGE project \citep[debuted in][]{2020MNRAS.491.1656A}, modelled using the adaptive mesh refinement (AMR) tool {\sc ramses} \citep{2002A&A...385..337T}. Our cosmological parameters are based on data from the Planck satellite \citep{2014A&A...571A..16P}, and are $\Omega_{\rm m}=0.309$, $\Omega_{\rm \Lambda}=0.691$, $\Omega_{\rm b}=0.045$ and $H_0=67.77\,\text{km\,s}^{-1}\,\text{Mpc}^{-1}$. Our baryonic physics recipe, zoom-halo selection criteria, and our simulation methods are all described in detail in \citet{2020MNRAS.491.1656A}. For completeness, we briefly summarise them here. \par

The fiducial baryonic physics model includes Schmidt-law star formation \citep{1959ApJ...129..243S}, stellar evolution and stellar feedback budgets adopted from \citet{2013ApJ...770...25A}. Our feedback scheme models the injection of energy, momentum and metals from supernovae (SNe) into the interstellar medium. A time-dependent, uniform ultraviolet background is used to model the epoch of reionisation (based on \citealt{1996ApJ...461...20H} with modifications as described in \citealt{2020MNRAS.497.1508R}). The hydrodynamic grid spatial resolution approaches $\sim 3$\,pc and $\sim 20\,\mathrm{M}_{\odot}$ within the zoom region, allowing us to resolve the impact of individual SNe on the intergalactic medium \citep{2015MNRAS.451.2900K, Wheeler2019, 2020MNRAS.491.1656A}. \par

All simulations are initialised with the {\sc GenetIC} code \citep{2021ApJS..252...28S} which allows us to forensically explore the impact of assembly history on galaxy properties using the genetic modification (GM) approach \citep{2016MNRAS.455..974R, 2018MNRAS.474...45R}. This involves altering the initial conditions of a cosmological simulation in a way that maximises the chance that those alterations arise from a random Gaussian draw from a $\Lambda$CDM cosmology. \par

Detailed analysis was performed using the {\sc Tangos} database package \citep{tangos} and the {\sc Pynbody} analysis package \citep{pynbody}. We analyse four haloes previously described in \citet{2019ApJ...886L...3R} and \citet{orkney}, all of which have a final stellar mass of $\mathcal{O}(10^5)\,\text{M}_{\odot}$ and are selected from the EDGE suite to be most similar to the observed properties of EriII (see Table \ref{tab:eri_II_comparison}). Halo1459 GM:Later and Halo1459 GM:Latest are based upon the initial conditions of Halo1459, but have been modified with our GM treatment. Halo1459 GM:Later (Halo1459 GM:Latest) has had its formation delayed such that it is a third (a half) times less massive than Halo1459 at the time of reionisation. \par

We find that bound ensembles of stellar particles, which we tentatively identify as SCs, form naturally within all of our EDGE simulations. We will investigate the formation physics and properties of these SCs in forthcoming work (Taylor et al. 2022 in prep.). Here, we note that the long-term dynamical evolution of these SCs cannot be relied upon because the stars are not individually resolved, and the intra-particle forces are artificially damped by the $\sim 3$\,pc force-softening. Therefore, SCs that develop in EDGE (and other similar cosmological simulations) are prone to artificial dissolution. Furthermore, our small sample of SCs in EDGE is not sufficient to predict the probability distribution function of their initial sizes, masses and orbits. For these reasons, we model the SCs instead with follow-up simulations using an accurate direction summation code, {\sc nbody6df}. We describe these follow-up simulations, next. \par

\subsection{Direct summation \textit{N}-body simulations} \label{n-body simulations}

\begin{table*}
\begin{tabular}{lc|c|c|c|cl} 
\toprule
\textbf{$N$} & \textbf{M$_{\mathrm{ini}}$ [$\text{M}_{\odot}$]} & \textbf{$R_{\mathrm{halfmass, ini}}$ [pc]} & \textbf{$R_{\rm g, ini}$ [pc]} & \textbf{$v_{\mathrm{ini}}/v_{\mathrm{circ}}$} & \textbf{Host $\gamma$} & Description \\
\midrule
$2^{16}$ & $4.18\times10^4$ & 2.5 & 45 & 1 & 1 & Halo1459 DMO: pristine $\Lambda$CDM cusp\\
$2^{15}$ & $2.08\times10^4$ & 5 & 100 & 0.75 & 0.75 & \textcolor{Halo1}{Halo1445}: weakened cusp\\
$2^{14}$ & $1.00\times10^4$ & 10 & 200 & 0.50 & 0 & \textcolor{Halo2}{Halo1459}: partial core\\
$2^{13}$ & $5.14\times10^3$ & 15 &  & 0.25 & 0 & \textcolor{GM}{Halo1459 GM:Later}: partial, time-evolving core\\
 & & & & & 0 & \textcolor{Fcore}{Fcore}: full core\\
\bottomrule
\end{tabular}
\caption{Overview of the parameter grid used for the {\sc nbody6df} simulation suite. From left to right, the columns give: the initial star numbers in the cluster and the corresponding total mass; the initial 3D half-mass radius used to initialise the cluster; the initial galactocentric distance of the cluster orbit; the initial tangential orbital velocity; and the central slope parameter $\gamma$ of the Dehnen model used for the host potential (where $\gamma=0$ is a core and $\gamma=1$ is a cusp), alongside a description of the potential. In total, this table describes 960 unique simulations.}
\label{tab:nbody_sims}
\end{table*}

We simulate the evolution of SCs in the EDGE dwarfs using a variant of the {\sc nbody6} code \citep{1999PASP..111.1333A} -- a Graphics Processing Unit (GPU) enabled direct $N$-body simulation tool \citep{2012MNRAS.424..545N}, which uses regularisation to model multiple-order stellar encounters. Stellar evolution is implemented with the standard `Eggleton, Tout and Hurley' option in {\sc nbody6} \citep{1989ApJ...347..998E, 1997MNRAS.291..732T, 2000MNRAS.315..543H, 2002MNRAS.329..897H, 2008LNP...760..283H}. \par

Our chosen variant is {\sc nbody6df} \footnote{\url{https://github.com/JamesAPetts/NBODY6df}} \citep*{2015MNRAS.454.3778P, 2016MNRAS.463..858P}, which introduces the effects of dynamical friction \citep[first described in][]{1943ApJ....97..255C}. Dynamical friction is a drag force which acts on a body as it passes through a background of lighter bodies, and would normally manifest as a result of interactions between a SC and background stars, gas and dark matter. {\sc nbody6df} uses a semi-analytic approach to calculate the orbital decay of a SC due to dynamical friction against the background, eliminating the need to simulate the full galactic context. The code relies upon a distribution function to calculate the dynamical friction force, which is analytic for a Dehnen potential \citep{1993MNRAS.265..250D}:
\noindent\begin{equation}
\label{dehnen.eq}
\rho_{\rm Dehnen}(r) = \rho_{0} \left ( \frac{r}{r_{\rm s}} \right )^{-\gamma}  \left (1+\frac{r}{r_{\rm s}} \right )^{\gamma-4},
\end{equation}
where $\rho_{0}$ is the central density, $r_{\rm s}$ is the scale radius and $\gamma$ is a variable used to set the logarithmic slope of the inner density profile ($\gamma = 0$ corresponds to a core and $\gamma = 1$ corresponds to a cusp). A range of options for the variable $\gamma$ are already implemented in {\sc nbody6df} based on the analytic solutions from \citet{1993MNRAS.265..250D}. As such, we make use of these by fitting the Dehnen profiles to the EDGE dwarf density profiles, as described in \S\ref{ramses_results_2}.\par

Critically, the semi-analytic dynamical friction model in {\sc nbody6df} is able to reproduce the `core-stalling' phenomenon whereby dynamical friction ceases when the orbital radius of the SC approaches its tidal radius \citep{2006MNRAS.373.1451R,2006MNRAS.368.1073G,2010ApJ...725.1707G, 2015MNRAS.454.3778P,2016MNRAS.463..858P,2018ApJ...868..134K,2021arXiv210305004B}. For a Dehnen potential, this radius can be estimated as:
\noindent\begin{equation}
\label{corestalling.eq}
r_{\mathrm{stall}} = \left [ \frac{M_{\mathrm{SC}}}{M_{\mathrm{g}}} \left ( r_{\mathrm{s}}^{2-\gamma} + r_{\mathrm{s}}\gamma \right ) \right ]^{1/(3-\gamma)},
\end{equation}
where $M_{\mathrm{SC}}/M_{\mathrm{g}}$ is the ratio of the SC mass to the host galaxy mass. \par 

The initial conditions for each SC were constructed with the tool {\sc McLuster} \citep{2011MNRAS.417.2300K} using a spherical Plummer density model \citep{1911MNRAS..71..460P} and a Kroupa IMF \citep{2002Sci...295...82K, 2010MNRAS.401..105K} with stellar masses in the range $0.1 \leqslant \mathrm{M}_{\odot} \leqslant 100$. This corresponds to a mean stellar mass of $\approx 0.63\,\text{M}_{\odot}$. We use an initial mean metallicity of $1\times 10^{-4}$, the lowest available in native {\sc nbody6}. We assume zero primordial binaries and that stars which evolve into black holes have no natal kick velocity. Both of these choices remove potential sources of dynamical heating and so maximise SC survivability. We retain unbound stars in the full integration of the system as we find that mass-loss rates are slightly reduced. All our SCs survive longer than a Hubble time when simulated in isolation. \par

We create initial conditions for SCs over a grid of masses (star numbers: $N=2^{13}, 2^{14}, 2^{15}, 2^{16}$, yielding the initial masses given in Table \ref{tab:nbody_sims}), initial half-mass radii ($ R_{1/2}/\text{pc}=2.5, 5, 10, 15$), orbital radii ($R_{\rm g}/\text{pc}=45, 100, 200$), and orbital eccentricities ($v_{\mathrm{ini}}/v_{\mathrm{circ}}=0.25, 0.50, 0.75, 1.00$), where $v_{\rm circ}$ is the circular velocity in a Dehnen potential:
\noindent\begin{equation}
\label{circ.eq}
v_{\mathrm{circ}}^2(r) = \frac{G M_{\rm g} r^{2-\gamma}}{(r + r_{\rm s})^{3-\gamma}}.
\end{equation}
The combinations of initial parameters are described in Table \ref{tab:nbody_sims}. Each permutation was then inserted into a background potential described by spherically symmetric Dehnen profile fits, which are discussed further in \S\ref{ramses_results_2}. These profile fits were chosen to represent a variety of central density slopes. Each simulation in the suite was run for either a Hubble time (13.8\,Gyr) or until SC dissolution, which was determined as the time when the number of bound cluster stars drops below 10.\par

The centre of each SC was determined using the mass-weighted shrinking spheres method of \citet{2003MNRAS.338...14P}, limited to stars with masses $\leqslant 10\,\text{M}_{\odot}$. All luminosities were converted to V-band with bolometric corrections based on fits from \citet{1998JRASC..92...36R}, after which we performed a cut on stars with luminosities $0.1 \geqslant L_{\rm V}/\text{L}_{\rm \odot,V} \geqslant 100$. This is because overly bright stars are normally removed from observational analysis, and faint stars escape detection. Half-light radii were calculated with bootstrapping analysis based on the total bound V-band luminosity. Black holes and compact remnants were excluded from our analysis when calculating optically derived values, such as the half-light radius. \par

Note that, although it is most similar to EriII's observed properties (see Table \ref{tab:eri_II_comparison}), we do not consider SC survival within Halo1459 GM:Latest. This is because, due to a late, low stellar mass, cuspy merger, its final density profile ends up being rather cuspy (see the discussion of this in \citealt{orkney}). As such, its behaviour in terms of SC survival is already covered by Halo1445. Nonetheless, it would be interesting to explore this galaxy further in future work. It has a half-light radius and stellar content similar to EriII, precisely because it assembles late from many minor mergers (see \citet{2019ApJ...886L...3R}). These minor mergers themselves may interact with the central SC, potentially explaining some of its puzzling properties. Exploring this is beyond the scope of this present work.

\section{Results} \label{Results}

\subsection{Cosmological simulations} \label{cosmo}

\subsubsection{EDGE ultra-faint dwarf candidates}\label{sec:EDGE-Eri}

The key observational properties of our EDGE UFDs are presented in Table \ref{tab:eri_II_comparison}. For comparison, we include properties for the UFD EriII, with values taken from the literature. The chosen EDGE galaxies are selected because they compare best to EriII in terms of their virial mass, stellar mass and their status as isolated reionisation fossils, but they are not intentionally designed to resemble EriII. Indeed, the EDGE stellar masses are somewhat greater than the upper limit measured for EriII. Notice that for Halo1459, the stellar mass and stellar metallicity drop as formation time is delayed (see GM:Later and GM:Latest). A delayed formation time also leads to a larger half-light radius and stellar velocity dispersion. These trends are a direct result of the assembly history and are described fully in \citet{2019ApJ...886L...3R}. The later forming versions of Halo1459 approach more and more closely the observed properties of EriII in just about every respect -- $M_*$, $R_{1/2}$, [Fe/H], $\sigma_{\rm LOS}$ and even the dynamical mass within $R_{1/2}$. This suggests that EriII is likely a late-former, following the arguments in \citet{2019ApJ...886L...3R}, which could be key to understanding its puzzling SC.

In our EDGE simulations, galaxies that form later are larger and have lower metallicity because their stellar content comprises a larger fraction of accreted stars. This is why, as we genetically engineer Halo1459 to form later and later, its size grows and its metallicity drops \citep{2019ApJ...886L...3R}. Note, however, that forming late is not sufficient on its own. Halo1445 forms even later than Halo1459 GM:Latest, yet its half stellar mass radius is a relatively compact $100\,$pc. This occurs because Halo1445 forms late but from fragments that have little/no stellar mass. Hence, its resulting size remains compact, while its metallicity is not as low as Halo1459 GM:Latest. For the above reasons, the galaxies drawn from EDGE that are most similar to EriII are those that form late and whose stellar populations comprise a high fraction of accreted stars, like Halo1459 GM:Latest. We expect these trends would be replicated across different simulations, although alternative galaxy-formation models may find that EriII is better-matched by galaxies with a median assembly history, with late assemblers being even larger and more metal poor. \par

\subsubsection{Mass assembly and star formation} \label{ramses_results_1}

\begin{figure}
\includegraphics[trim={0cm 1cm 0cm 0cm}, clip=False, width=\columnwidth]{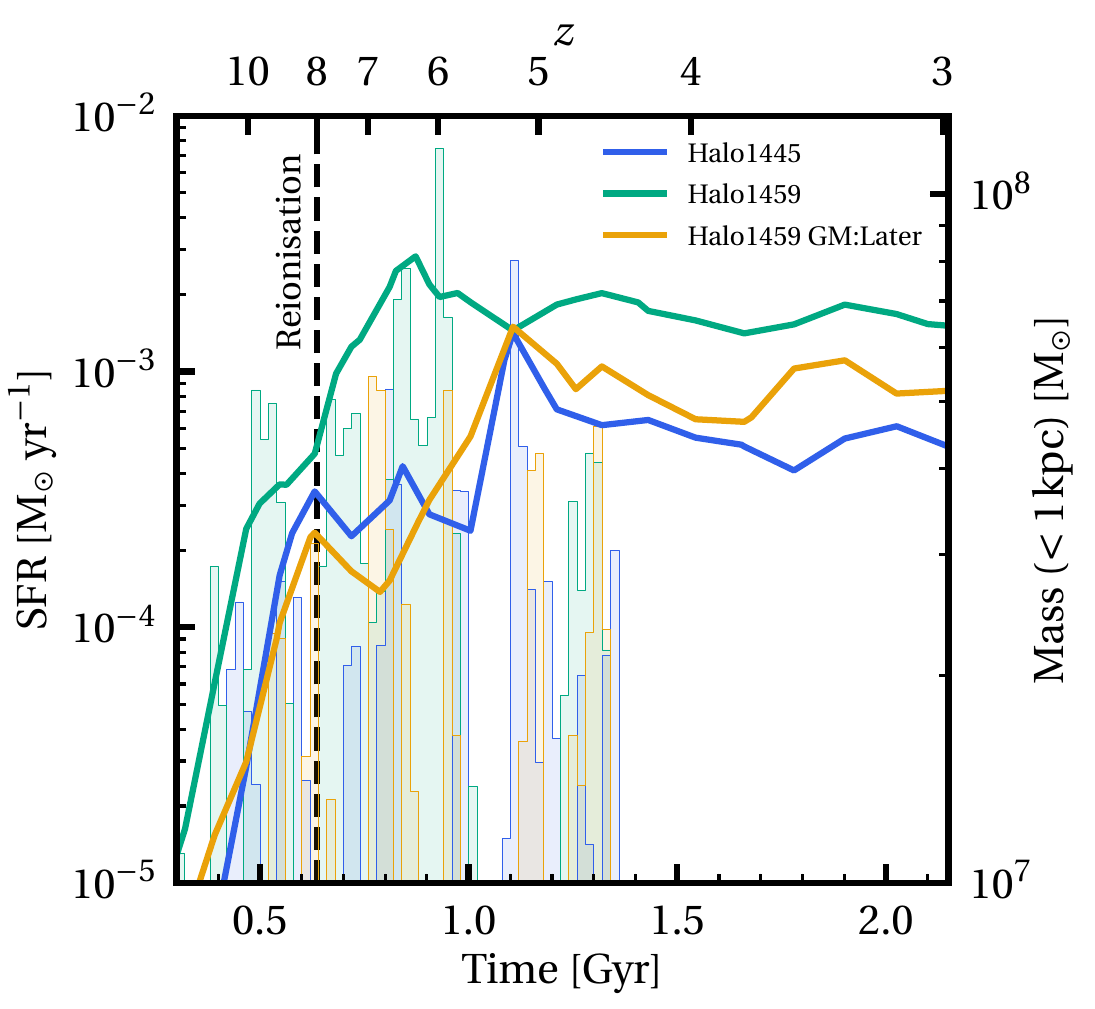} \\
\caption{The star formation rate (left-hand y-axis, histogram) and mass assembly history (right-hand y-axis, lines) for our EDGE UFD simulations, at early times. The star formation rate is averaged over bin sizes of 0.02\,Gyr and takes into account all stars formed within the virial radius $R_{\rm 200c}$ of the main progenitor halo. The mass assembly history considers the total mass contained within the inner 1\,kpc of the main progenitor halo. A dashed line indicates the start of reionisation. The dwarfs continue to form stars until their remaining cold gas is depleted \citep[see][]{2020MNRAS.497.1508R}. Only more massive EDGE dwarfs (not included in this paper) are able to accrete sufficient gas after reionisation to reignite star formation.} \label{fig:SFH_mass}
\end{figure}

Our EDGE haloes attain $M_{\rm 200c}$ halo masses of approximately $1.5\times10^9\,\mathrm{M}_{\odot}$ by $z=0$, where we define $M_{\rm 200c}$ as the mass within a spherical volume encompassing 200 times the critical density of the universe. All haloes form stars when baryonic physics is switched on, although this star formation is confined to early times ($z>4$). Star formation continues for some time after the onset of reionization, but is quenched after the remaining cool gas within the halo is exhausted (as shown in \citealp{2020MNRAS.497.1508R}, and see also \citealt{2004ApJ...600....1S, 2015MNRAS.454.2092O}).\par

The star formation (left-hand axis) and mass growth histories (right-hand axis) are shown in Figure \ref{fig:SFH_mass} up to $z=3$. The mass growth is defined as the mass enclosed within 1\,kpc rather than the virial mass $M_{\rm 200c}$, because the central mass is what is most important for the survival of low-orbit SCs, and because $M_{\rm 200c}$ undergoes a pseudo-growth due to the lowering background density of the Universe as it expands \citep{2007ApJ...667..859D, 2013ApJ...766...25D}. The mass enclosed within 1\,kpc remains roughly constant after 1\,Gyr, despite the continued accretion of dark matter. The star formation in all haloes is bursty, which is a necessary quality in driving the gaseous flows responsible for dark matter core formation \citep{2012MNRAS.421.3464P}. Indeed, we observe a fluctuating central gas mass in all haloes at early times \citep{orkney}. Halo1459 assembles its central mass most rapidly and has the highest final central mass, along with the most substantial star formation. The larger dynamical potential well aids in the compression of gas into star-forming clouds. This is a consequence of the well-established concentration-formation relation \citep[e.g.][]{2002ApJ...568...52W, 2014MNRAS.441..378L, 2020MNRAS.498.4450W}. \par

\subsubsection{Density profiles} \label{ramses_results_2}

\begin{figure}
\centering
\setlength\tabcolsep{2pt}%
\includegraphics[ trim={0cm 1cm 0cm 0cm}, clip=False, width=\columnwidth, keepaspectratio]{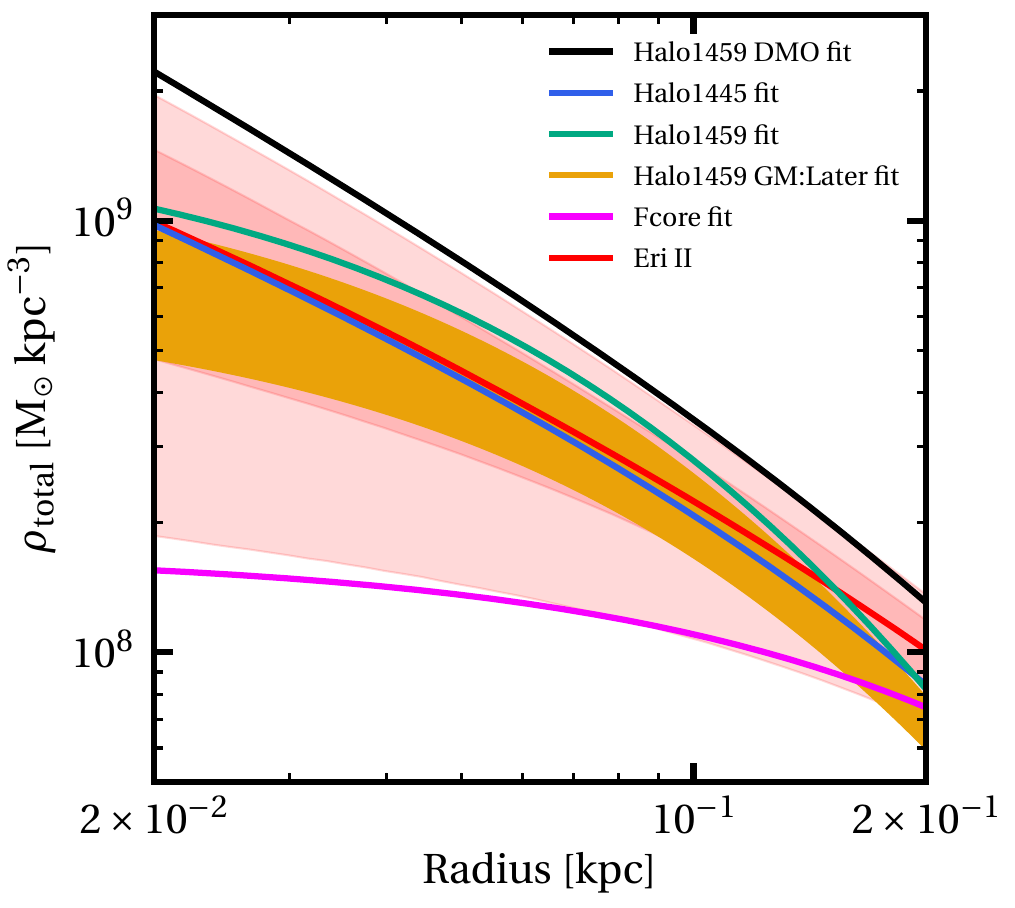}
\caption{Fits to the total inner 3D density profiles of EDGE haloes at $z=3.2$. The fits to Halo1459 GM:Later are shown as a filled region, with the upper (lower) limit representing the fit at $z=4$ ($z=0$). We also show constraints on the density profile of EriII based on the `core+tides' profile fit from \citet{2021arXiv211209374Z}, where the dark and light shaded bands correspond to the 68 percent and 95 percent confidence limits, respectively.} \label{fig:Dehnen_fit}
\end{figure}

In Figure \ref{fig:Dehnen_fit}, we present spherically symmetric Dehnen profile fits to the \textit{total} mass distribution of our EDGE haloes after the cessation of star formation. The raw density profiles of the dark matter distribution were reported previously in \citet{orkney}. For Halo1445 and Halo1459, we fit the halo density profiles at $z=3.2$, by which time the inner halo is fully assembled and stable until $z=0$ (see Figure \ref{fig:SFH_mass}). For Halo1459 GM:Later, which experiences additional late time heating from minor mergers, we make a fit both at $z=4$ and $z=0$, interpolating between them to obtain a time varying potential that matches the simulation data at all times (using the method described in \citealt{2019MNRAS.488.2977O}). We also produce a fit for Halo1459 DMO at $z=3.2$ to provide a $\gamma=1$ cusp reference, but do not fit profiles to the other DM-only simulations. These fits are produced with the python module {\sc LmFit} \citep{newville_matthew_2014_11813}, considering only the portion of the profile that is unaffected by numerical relaxation (as discussed in \citealt{orkney}, Appendix A). Our fits accurately reproduce the simulation density over the ranges shown, but deviate from the data in the outer regions because a Dehnen profile density falls off as $r^{-4}$ in its outskirts whereas our simulated haloes fall off as $r^{-3}$. These fits are, nonetheless, suitable for our purpose here, since we limit our analysis to initial orbital radii of 200\,pc or less. \par

Included in Figure \ref{fig:Dehnen_fit} is the cored Dehnen profile from \citet{2018MNRAS.476.3124C}. Those authors found that SCs evolving in this profile were best able to match the properties of the SC in EriII. Hereafter we refer to this profile as `Fcore'; it has a substantially lower density core than any of our EDGE haloes. We also show observational constraints on the dark matter density profile of EriII in red, based upon the `core+tides' fit performed by \citet{2021arXiv211209374Z}, assuming a CDM cosmology. This fit is made with pyGravSphere \citep{2017MNRAS.471.4541R} using stellar line-of-sight velocities as measured by the MUSE-Faint survey. The shaded bands correspond to 68 percent and 95 percent confidence limits. The Dehnen profiles fit to our EDGE dwarfs are mostly contained within the 68 percent limits, and the Fcore profile sits at the edge of the 95 percent limits. Therefore, the profiles we consider are well bracketed by observational fits. \par

\subsection{\textit{N}-body star cluster simulations} \label{nbody}

\subsubsection{SC properties after a Hubble time}\label{sec:SCprops}

\begin{figure}
\centering
\setlength\tabcolsep{2pt}%
\includegraphics[ trim={0cm 1cm 0cm 0cm}, clip=False, width=\columnwidth, keepaspectratio]{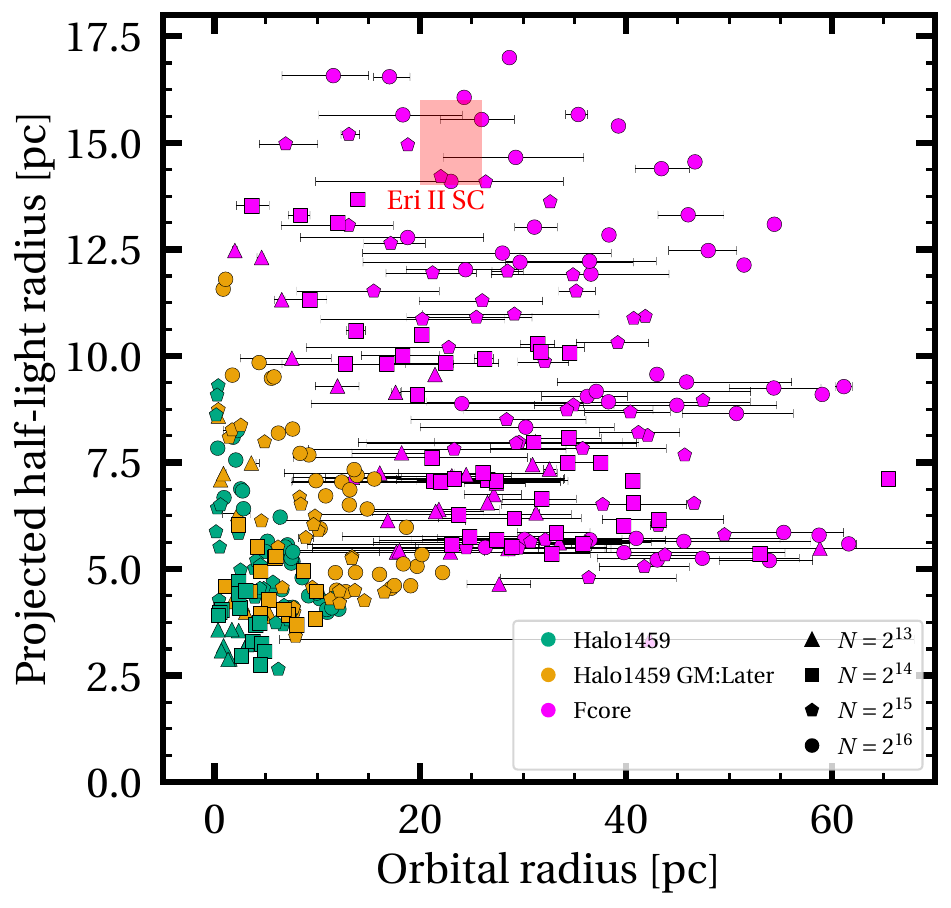}
\caption{The projected half-light radius versus the orbital radius of all surviving SCs from our simulation suite after a Hubble time. The marker style indicates the initial number of stars, $N$, in each SC, while the colour indicates the host mass profile model, as marked in the legend. Each point represents the average projected half-light radius of the current SC orbit, with the `error bars' marking the orbital apocentre and pericentre. Also included is a red box marking the properties of the SC in EriII \citet{2020arXiv201200043S}. No SCs from Halo1445 or Halo1459 DMO survive after a Hubble time and so these two host potentials do not appear on the Figure. A small number of SCs in the Fcore potential lie beyond the 70\,pc limit of this plot.} \label{fig:final_parameters}
\end{figure}

We find that only those SCs hosted by cored potentials (models: Halo1459, Halo1459 GM:Later and Fcore) survive for a full Hubble time. Just two SCs in a weakened cusp potential (Halo1445) survive longer than 6\,Gyr, while only three SCs in the pristine cusp potential (Halo1459 DMO) survive longer than 3\,Gyr. Therefore, we confirm the results of prior work that SC survival is extremely sensitive to the inner potential slope of the host \citep[e.g.][]{2017ApJ...844...64A,2018MNRAS.476.3124C,2019MNRAS.488.2977O}.

The projected half-light and orbital radius of all surviving SCs are presented in Figure \ref{fig:final_parameters}. The values are averaged over the last pericentre-apocentre passage, where the most recent pericentre and apocentre are marked by the horizontal `error bars'. The observed properties of the SC in EriII, with uncertainties, are indicated by a red box. \par

The orbital response to dynamical friction is so strong that the majority of SCs approach their core-stalling limit within a few Gyrs. However, some SCs in the Fcore potential have yet to reach this limit by a Hubble time. The orbital radius then continues to decay as the SC is stripped of mass, as predicted by Equation \ref{corestalling.eq}. SCs inhabiting the Fcore potential adopt a wider range of final orbital radii depending on their initial orbits and retain a larger degree of their initial orbital eccentricity, whereas the orbits of SCs in other host potentials are rapidly circularised. Despite a stronger dynamical friction force, the most massive SCs maintain slightly higher final orbital radii due to their larger core-stalling radius.\par

As found in prior work, SCs within the Fcore model grow to significantly larger sizes due to the low background potential and reduced tidal stripping \citep{2018MNRAS.476.3124C}. Amongst these, the very largest SCs are those with the largest $R_{\rm halfmass,ini}$, with $R_{\rm halfmass,ini}=15\,$pc bracketing the size of the SC in EriII. \par

\begin{figure}
\centering
\setlength\tabcolsep{2pt}%
\includegraphics[ trim={0cm 1cm 0cm 0cm}, clip=False, width=\columnwidth, keepaspectratio]{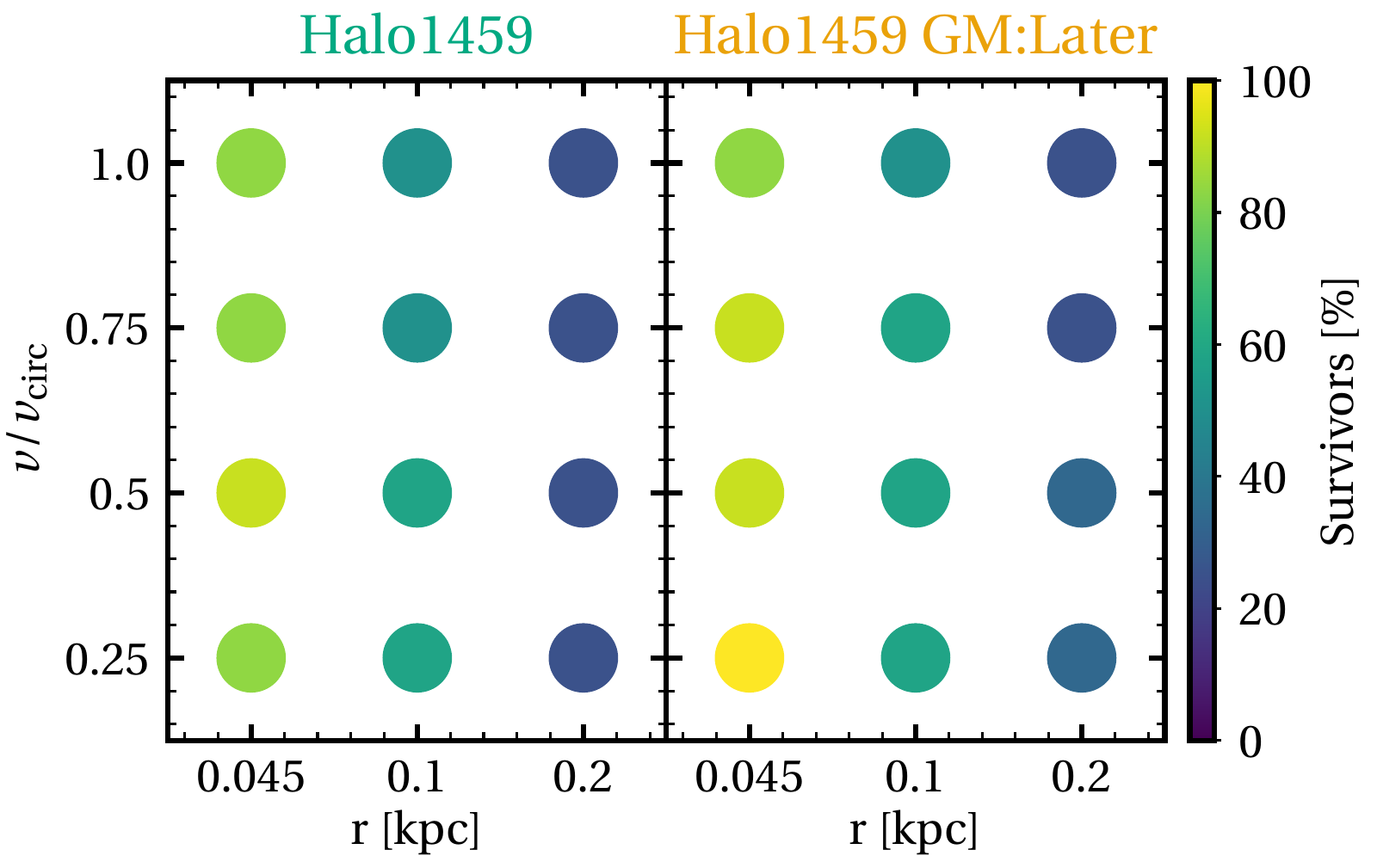}
\caption{The Hubble time survival of SCs in two host potentials. Each circle represents the survival percentage of SCs in Halo1459 and Halo1459 GM:Later, partitioned by initial velocity and orbital radius. We have omitted SCs with $R_{\rm halfmass, ini}=15\,\text{pc}$ because these rarely survive, regardless of their initial orbital parameters.} \label{fig:survivorship}
\end{figure}

Following the premise that harsher gravitational tides act to destroy orbiting SCs more rapidly, it is natural to assume that there are simple relationships between SC survival and the initial orbit. Normally, it is expected that more eccentric orbits, which plunge the SC in and out of a stronger background potential, result in shorter-lived SCs due to the impact of tidal shocks \citep{1994AJ....108.1403W, 1995ApJ...438..702K, 1999ApJ...514..109G, 2003MNRAS.340..227B, 2019MNRAS.486.5879W}. To explore this, in Figure \ref{fig:survivorship} we compare the survival of a selection of SCs in the Halo1459 and Halo1459 GM:Later host potentials, grouped by initial orbital velocity and radius. As expected, survival is improved with lower initial orbital radii. However, survival is largely independent of initial orbital eccentricity, with variation mostly below the level of Poisson noise. \par

Whilst it is generally the case that eccentric orbits encourage SC dissolution, it is also the case that dynamical friction is enhanced on eccentric orbits (which in turn acts to circularise the orbit). If the orbit of a SC decays to within the core region of the host potential, which typically occurs over Gyr time-scales, then it is better able to survive. In the case of the host potentials considered here, these two competing processes balance each other remarkably well. This is not the case for our steeper host potentials (models: Halo1445 and Halo1459 DMO), in which SCs survive the longest when on more circular orbits. \par

\subsubsection{SC shape after a Hubble time}\label{sec:hubbleshape}

\begin{figure}
\centering
\setlength\tabcolsep{2pt}%
\includegraphics[ trim={0cm 1cm 0cm 0cm}, clip=False, width=\columnwidth, keepaspectratio]{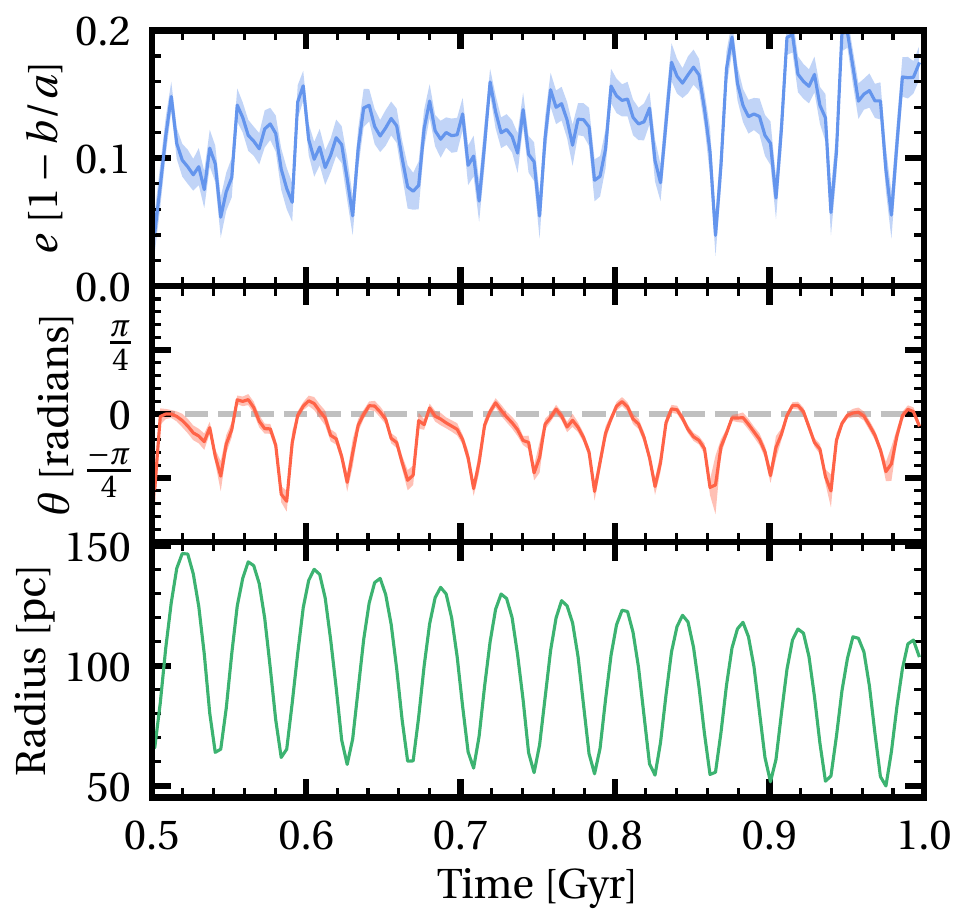}
\caption{The evolution of the SC shape for a single example simulation over 0.5 Gyr (from $0.5-1$\,Gyr in simulation time). The upper panel shows the ellipticity $e$, the middle panel shows the orientation $\theta$ (where $\theta=0$ is aligned with the galactic centre), and the lower panel shows the galactocentric orbital radius. The filled regions are the $1\sigma$ uncertainty determined from bootstrapping. For eccentric orbits, the SC shape and orientation is highly dependent on the position along the orbit.} \label{fig:shape_evolution}
\end{figure}

\begin{figure}
\centering
\setlength\tabcolsep{2pt}%
\includegraphics[ trim={0cm 1cm 1cm 0cm}, clip=False, width=\columnwidth, keepaspectratio]{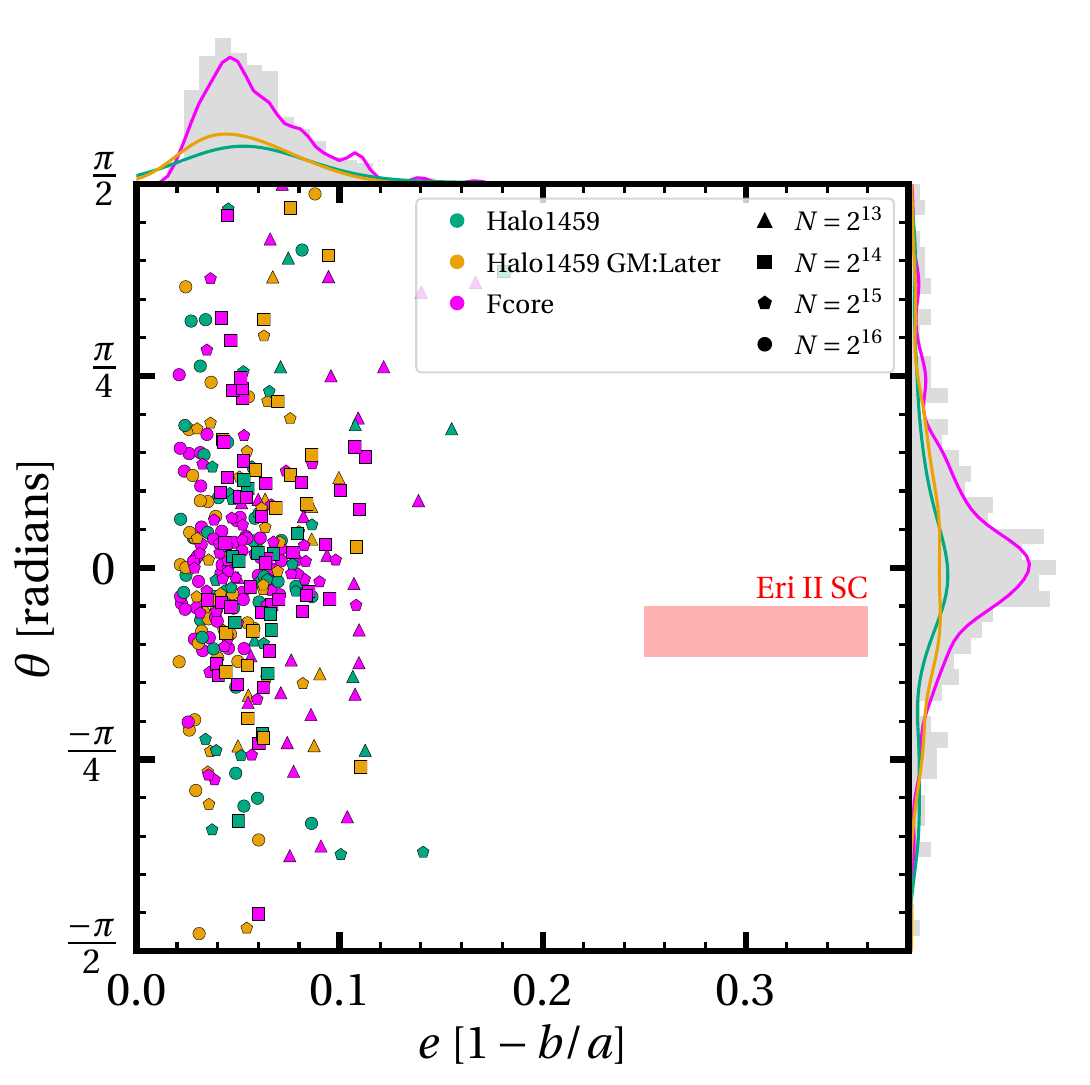}
\caption{The orientation $\theta$ versus ellipticity $e$ for all surviving SC simulations after a Hubble time, as viewed face-on to the orbital plane, calculated at the orbital apocentre (maximising $e$ and minimising $|\theta|$). Histograms, which include Gaussian kernel density estimations for each host profile, are shown at the borders. The red region indicates the observed properties of the SC in EriII from \citet{2020arXiv201200043S}.} \label{fig:shape_analysis}
\end{figure}

The ellipticity of the SC in EriII is $0.31\substack{+0.05 \\ -0.06}$ \citep{2020arXiv201200043S}, which is exceedingly high when compared to Milky Way Globular Clusters \citep{2010arXiv1012.3224H}. In this section we investigate the shapes of our simulated SCs in terms of their ellipticity and orientation. We define the ellipticity, $e$, as $1-b/a$, where $a$ and $b$ are the long-axis and short-axis of an ellipse fit to the SC stars. The orientation, $\theta$, is defined as the inclination in the long-axis of the SC ellipse with respect to the centre of the host potential (and is distinct from the `alignment' measurement reported in \citealt{2020arXiv201200043S}). Both values are calculated in projection, and are consistent with literature definitions. Our calculations are performed with the shape finder in {\sc pynbody}, modified for use with projected stellar distributions. We perform bootstrapping analysis over 1000 iterations for every shape fit, generally finding low uncertainties except for the degenerate orientation angles of near-circular SCs. We investigate the robustness of our method in Appendix \ref{appendix:a}, and show that it compares well to a typical observational method. \par

Throughout, we orient our SCs face-on to the orbital plane. This does not materially affect our results and conclusions. We explore this explicitly in Appendix \ref{app:edgeon}, where we consider the other extreme of orienting the SCs edge-on to the orbital plane. In this case, we find that tidally stripped debris along the line of sight can generate higher ellipticities (with a secondary sub-dominant effect caused by tidal compression). However, this occurs only for models in which the SCs are too small and not offset enough to be consistent with the SC in EriII. \par

In Figure \ref{fig:shape_evolution}, we show the evolution of the ellipticity $e$ and the orientation $\theta$ for one example SC over a Gyr. The initial properties of our chosen example are $M_{\rm ini}=2.08\times10^4\,\text{M}_{\odot}$, $R_{\rm halfmass,ini}=5\,\text{pc}$, $R_{\rm g,ini}=200\,\text{pc}$ and $v_{\rm ini}/v_{\rm circ}=0.50$, orbiting in the Halo1459 model. The SC undergoes total dissolution after 1.8\,Gyr. This example was chosen purely because it clearly exhibits periodic trends, but there are comparable patterns for other simulations on non-circular orbits. The figure shows that $e$ (upper panel) and $\theta$ (middle panel) vary periodically with the orbital radius (lower panel). $e$ is maximised at apocentre, at which time the cluster is most aligned with the galactic centre ($\theta=0$). The periodicity in $e$ is a consequence of the tidal material sloshing around the SC body, and is strongest for simulations where there is significant tidal material. There are also sub-dominant long-term evolutionary trends in all simulations. \par

We show $e$ and $\theta$ for all surviving SC simulations in Figure \ref{fig:shape_analysis}. Motivated by the results in Figure \ref{fig:shape_evolution}, we estimate the SC shape at the most recent orbital apocentre. The coloured Gaussian kernel density estimations show the contributions from the different host potential profiles. This shows that, whilst all SCs show a preference for aligning with the centres of their host galactic potentials, their ellipticity remains much lower than that of the SC in EriII. As anticipated in \S\ref{sec:intro}, this is because all of the surviving SCs orbit close to the centre of $\gamma = 0$ background potentials and, therefore, experience little tidal perturbation. We explore the implications of this in Section \ref{discussion}. \par

\subsubsection{Primordial ellipticity} \label{primordial_e}

\begin{figure}
\centering
\setlength\tabcolsep{2pt}%
\includegraphics[ trim={0cm 1cm 0cm 0cm}, clip=False, width=\columnwidth, keepaspectratio]{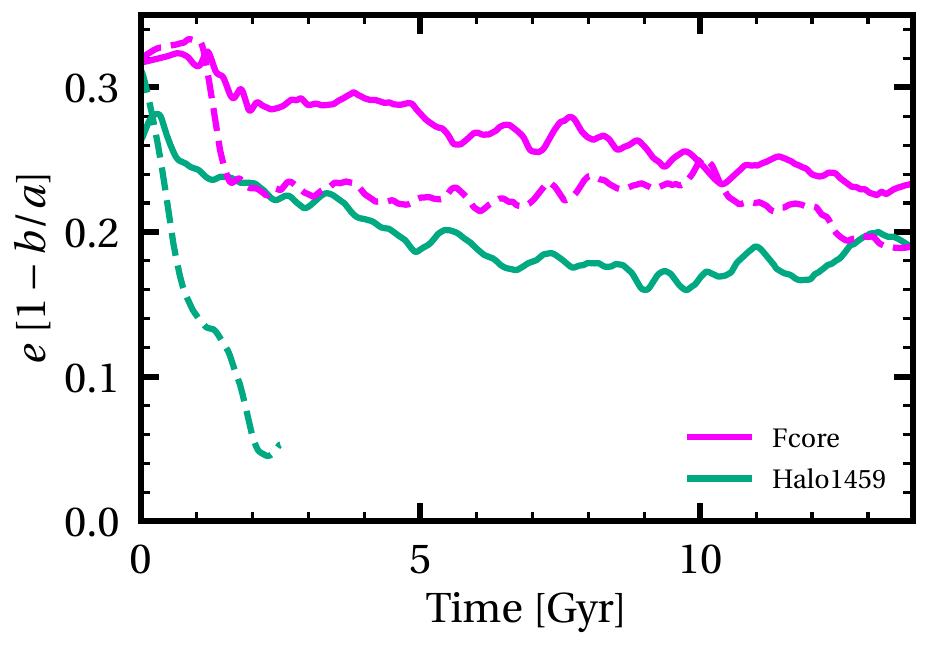}
\caption{Evolution of the projected ellipticity for initially rotating and elliptical SCs in a partial core (Halo1459; green) and full core (Fcore; magenta) background potential. The data have been smoothed to remove periodic trends. The solid (dashed) lines represent orbits starting at a radius of $45\,$pc ($200\,$pc). The SC in the `Halo1459' potential at an initial orbital radius of $200\,$pc dies shortly after 2\,Gyr.} \label{fig:primordial_e}
\end{figure}

In \S\ref{sec:SCprops} and \S\ref{sec:hubbleshape}, we have shown that SCs survive for long times in all cored potentials, while only in the shallower Fcore potential do they reach a size and offset comparable to EriII's SC. However, due to the lack of tidal forces within the core, these SCs retain their initial sphericity and fail to match the high ellipticity of EriII's SC. This appears to present a significant problem for the full core scenario since prior work suggests that any initial ellipticity in EriII's SC should be rapidly erased, given its short relaxation time (see \S\ref{sec:intro} and \citealt{1958SvA.....2...22A,1976ApJ...210..757S, 1996A&A...308..453L,einsel99,bianchini13}).

In this section, we examine the above concern in more detail. Firstly, we note that young SCs are typically born elliptical due to rotational flattening, pressure anisotropies and/or SC-SC mergers \citep[e.g.][]{1983AJ.....88.1626F, 2018MNRAS.473.5591K, 2018ApJ...860...50F, 2018MNRAS.481.2125B}, with `primordial' birth ellipticities often higher than $e > 0.3$ \citep{getman18}. Pressure anisotropies, which arise when systems form from the violent relaxation of an aspherical collapse, are lost on relaxation timescales \citep{1978MNRAS.185..227A}. Ellipticity due to rotational flattening should also be reduced on a relaxation time-scale as two-body scattering moves high angular momentum stars (which are responsible for the elliptical shape) to the outskirts of the cluster, where they are lost to tides \citep{1999MNRAS.302...81E, 2017MNRAS.469..683T}. However, the tidal forces vanish inside a constant density core and there is no mechanism to carry away the high angular momentum material. This suggests that in a weak tidal field, primordial ellipticity should be `locked in', surviving for long times \citep{1997MNRAS.286L..39G}.

To test the above idea, we deform one of our SCs and add a rotation curve about the deformation axis of the following form:
\noindent\begin{equation}
v_{\rm rot} = \frac{2V_{\rm peak}}{R_{\rm peak}} \frac{R}{1+(R/R_{\rm peak})^2},
\label{rotcurve.eq}
\end{equation}
where $V_{\rm peak}$ is the maximum velocity and $R_{\rm peak}$ is the radius of maximum velocity (as in \citealt{1967MNRAS.136..101L, 2013ApJ...762...65M}).
We ensure that virial equilibrium and the overall SC volume are maintained. Through this, we create a SC with $M_{\rm ini} = 2.08\times10^4\,\text{M}_{\odot}$, $R_{\rm halfmass, ini}=5$, $e_{\rm ini}=0.31$ and rotation with a peak of $1.75\,$km\,s$^{-1}$ at the half-mass radius. In Figure \ref{fig:primordial_e}, we show the evolution of the ellipticity of this SC when evolved in two different background potentials. In both cases, we initialise the SC on a circular orbit at two different initial orbital radii of $45\,$pc (solid lines) and $200\,$pc (dashed lines). A high ellipticity is maintained for over a Hubble time in the Fcore potential, and is rapidly lost in the Halo1459 potential when at an initial offset of $200\,$pc. This suggests that the high ellipticity of EriII's SC does not in fact present a problem for the Fcore model. In such a model, it must simply reflect the ellipticity of EriII's SC at birth, which would be consistent with that of young SCs \citep{getman18}. Given the much larger parameter space required to explore the ellipticity and orbit of EriII's SC, we leave a more thorough investigation of how ellipticity affects the survival and evolution of SCs to future work. \par

\subsubsection{Star clusters in ultra-faint dwarfs via recent accretion}

We now explore an alternative scenario for EriII's SC in which it has been recently accreted rather than surviving with its current properties for long times. Whilst our EDGE dwarfs assemble their central mass by $z=4$, there are still a significant number of mergers throughout their lives. Indeed, EriII is most similar to the very latest assembling EDGE dwarfs (see \S\ref{sec:EDGE-Eri} and Table \ref{tab:eri_II_comparison}). These mergers can contribute to late time dark matter heating as with Halo1459 GM:Later \citep{orkney}, but could also act as vehicles for depositing SCs. There are a total of five star-rich mergers which occur after $z=1$, however only one is massive enough to compare favourably to the SC in EriII. A SC that is donated via a subhalo merger would be protected from much of the destructive forces of gravitational tides until the dark matter envelope is stripped \citep[e.g.][]{2005ApJ...619..243M, 2005ApJ...619..258M, 2021arXiv210403635B, 2021arXiv211201265V}. The increased system mass would also cause the SC orbits to decay faster due to dynamical friction. Assuming that previously accreted SCs fall to the centre of EriII without yielding a detectable central SC (c.f. \citealt{2020arXiv201208058S}), it is possible that EriII experienced a long chain of such accretions of which only the most recent remains detectable. \par

\begin{figure}
\centering
\setlength\tabcolsep{2pt}%
\includegraphics[ trim={0cm 1cm 0cm 0cm}, clip=False, width=\columnwidth, keepaspectratio]{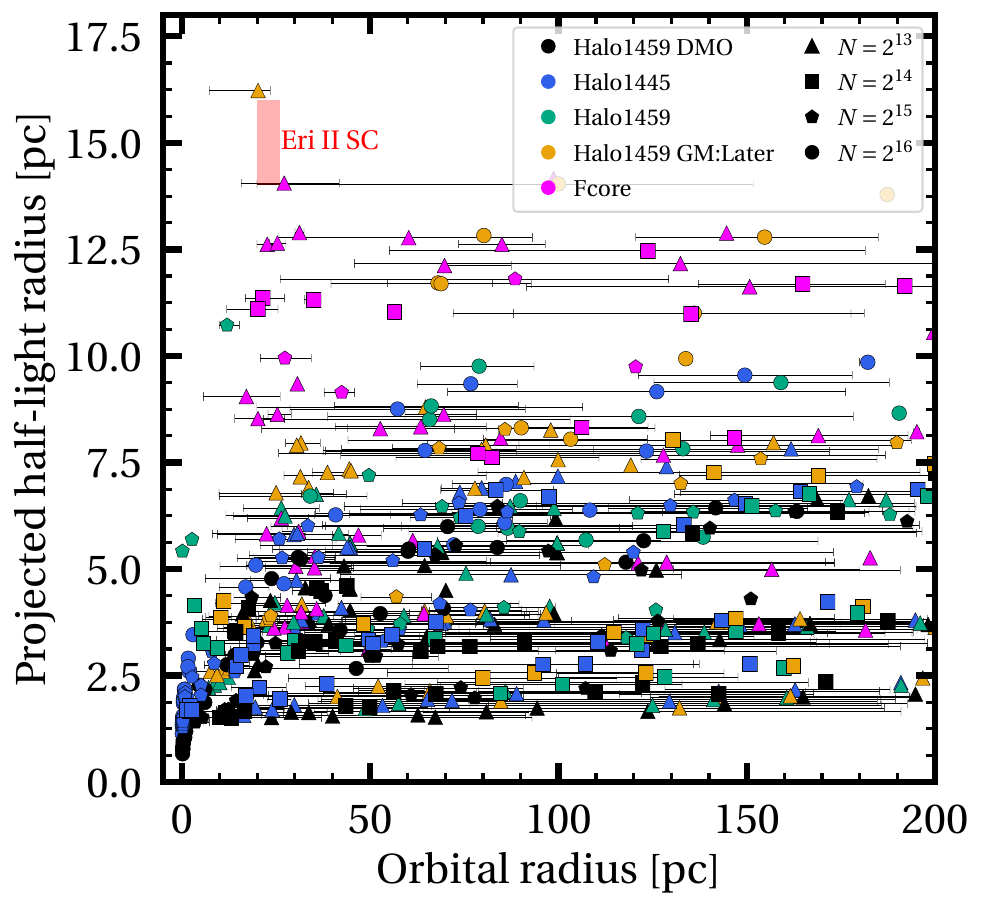}
\caption{The projected half-light radius versus the orbital radius of our SCs simulation suite, shown at the time in their evolution when they pass the lower mass estimate for the SC in EriII. The figure format is otherwise identical to that of Figure \ref{fig:final_parameters}.} \label{fig:dying_parameters}
\end{figure}

\begin{figure}
\centering
\setlength\tabcolsep{2pt}%
\includegraphics[ trim={0cm 1cm 1cm 0cm}, clip=False, width=\columnwidth, keepaspectratio]{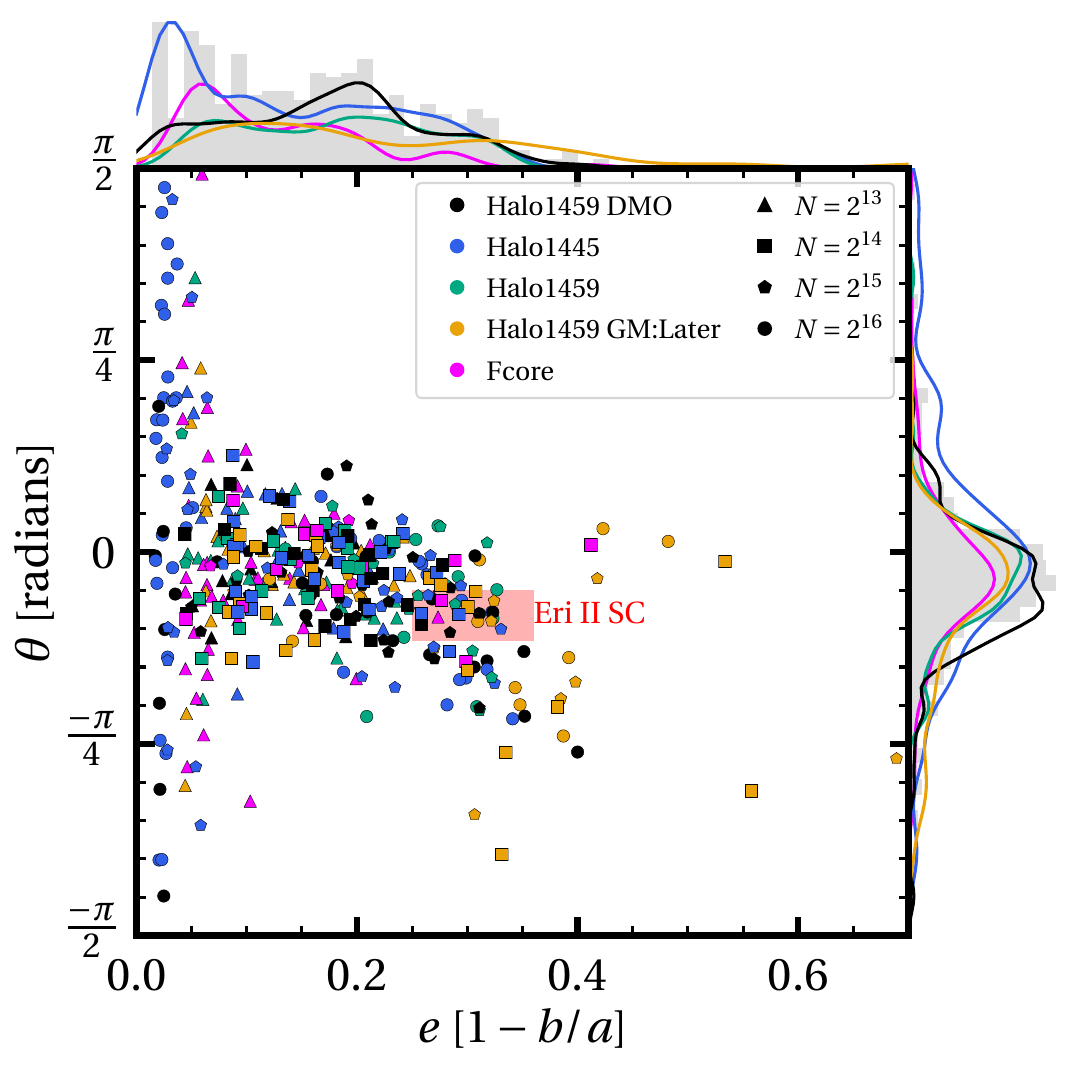}
\caption{The orientation $\theta$ versus ellipticity $e$ for our SCs simulation suite, shown at the time in their evolution when they pass the lower mass estimate for the SC in EriII. The figure format is otherwise identical to that of Figure \ref{fig:shape_analysis}.} \label{fig:dying_shape_analysis}
\end{figure}

So far, we have considered only SCs which have survived within their host potential for a Hubble time, but recently accreted SCs would not need to survive this long. Here, we instead consider SCs at a time when their bound mass drops just below the lower mass estimate of the SC in EriII (which we take to be $2.5\times10^3\,\text{M}_{\odot}$, following \citealt{2016ApJ...824L..14C, 2018MNRAS.476.3124C}), with the one additional constraint that they survive for at least 100\,Myr. Many of the SCs in this low-mass regime have been exposed to their host potentials for only a short time, and are close to total dissolution. In Figure \ref{fig:dying_parameters}, we once again show the projected half-light radius against orbital radius. Since we are now including SCs that are close to dissolution from tides, there are contributions from SCs in the steeper background potential models (models: Halo1459 DMO and Halo1445). Notice, however, that the half-light radii of the SCs in these steeper potentials never grow beyond $\sim10\,$pc and so they remain inconsistent with EriII's SC. In Figure \ref{fig:dying_shape_analysis}, we once again show $e$ and $\theta$. There is now a spray of SCs at higher $e$, predominantly formed by simulations in the densest potentials. There are also a small number of highly elliptical SCs in the Fcore potential, which are diffuse SCs with high $R_{\rm halfmass,ini}$ and are therefore more sensitive to gravitational tides. The spray is clustered around an orientation of $\theta=0$ at low $e$, but then begins to favour more misaligned $\theta$ at increasingly high $e$. This is because the presence of tidal tails contribute to the measured ellipticity, and these originate at the Lagrange points of the SC (which are always aligned with the galactic centre; e.g. \citealt{klimentowski09}). As the tidal tails grow longer they begin to trail and lead the SC orbit, which leads to a higher perceived $e$ and a larger orientation angle with respect to the galactic centre. We discuss the implications of this phenomenon, next. \par

\subsubsection{Star cluster tidal tails}

\begin{figure}
\centering
\setlength\tabcolsep{2pt}%
\includegraphics[ trim={0cm 1cm 0cm 0cm}, clip=False, width=\columnwidth, keepaspectratio]{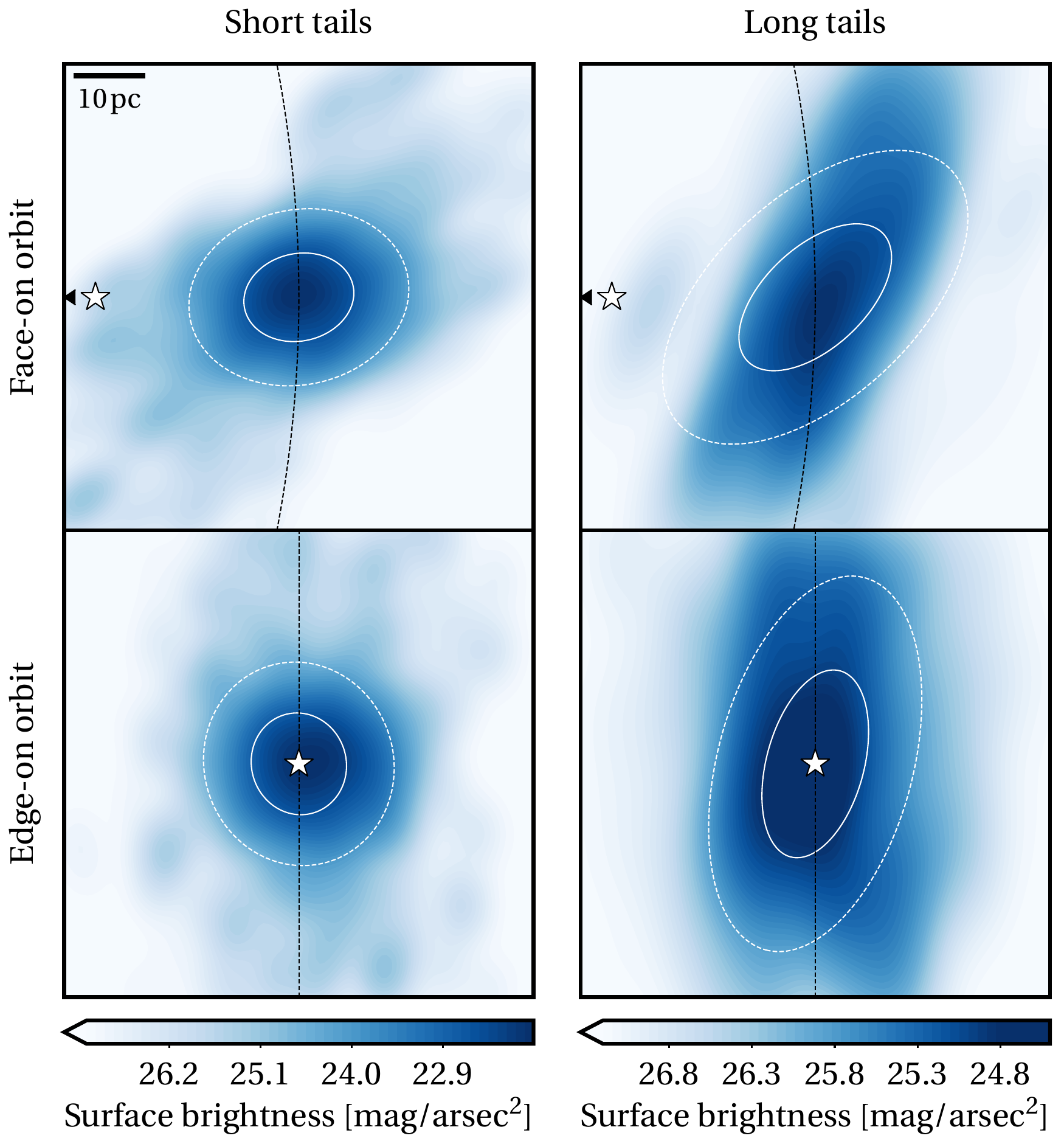}
\caption{Kernel density estimate plots of the V-band surface brightness for a SC simulation at two stages in its evolution, shown with two different orientations. A surface brightness background of 27.2 mag/arcsec$^2$ is included to mimic the central brightness of EriII \citep{2016ApJ...824L..14C}. The white star indicates the centre of the host potential, a black dashed line represents the SC orbit, and white solid/dashed lines represent a shape fit to the SC shown at $1\times$/$2\times$ the half-light radius. The left panels represent a younger SC with relatively short tidal tails coming from the Lagrange points. The right panels represent the SC on the verge of total dissolution, which has become smeared across its orbit.} \label{fig:tail_illustration}
\end{figure}

The SC ellipticities and orientations presented in Figures \ref{fig:shape_analysis} and \ref{fig:dying_shape_analysis} are as viewed in the plane of the SC orbit. However, the SC orbit in EriII may be viewed from an edge-on perspective, or somewhere in-between. This could change the perceived orientation and ellipticity. \par

We illustrate this key point in Figure \ref{fig:tail_illustration}, which shows an example SC simulation at orbital apocentre (thereby maximising its ellipticity).
The left panels show the SC at an early stage in its evolution. At this time, it is well aligned with the centre of the host potential and has only a low tidal tail brightness. The ellipticity ($e\approx0.2$) is clear when viewed face-on to the orbital plane (upper panel), but is diminished ($e\approx0.07$) when viewed edge-on to the orbital plane (lower panel). The right panels show the same SC in the final few Myrs before dissolution, at which stage it grows long tidal tails. These long tidal tails begin at the Lagrange points of the SC, then trail and lead the orbit. This leads to a shape fit that is highly elliptical ($e\approx0.48$) and misaligned with the centre of the host potential when viewed face-on to the orbital plane (upper panel). The SC now continues to appear highly elliptical ($e\approx0.53$) when viewed edge-on to the orbital plane (lower panel). \par
The presence of tidal tails, long or short, means that the apparent ellipticity will increase when considering larger cluster-centric radii, a behaviour that has been reported for EriII's SC (see the discussion in section 4 of \citealt{2020arXiv201200043S}). However, even our lowest mass SC simulations in this ``recent accretion'' scenario possess tails with surface brightness comparable to or exceeding the central brightness in EriII (27.2 mag/arcsec$^2$; \citealt{2016ApJ...824L..14C}) and should therefore be present in the observations of the galaxy. Yet, there is no sign of tidal material beyond two-times the half-light radius of the SC \citep[see figure 11,][]{2020arXiv201200043S}. This appears to rule out recently accreted SCs where the high ellipticity is due to significant tidal disruption.\par

\section{Discussion} \label{discussion}

\subsection{A full dark matter core in EriII}\label{largecore}

We have shown that a substantial dark matter core allows for the stable existence of large, elliptical and radially offset SCs, like that seen in EriII, over many Gyrs. Such a core would support both in-situ and accreted SCs. However, none of the UFDs in EDGE produce a central dark matter core of sufficient size and low enough density. We discuss next how such a large, low density, core might arise.

\subsubsection{Growing a full dark matter core} \label{coreforming}

\noindent
{\bf Full cores in $\Lambda$CDM $\mid$}

In \citet{orkney}, we showed that haloes with significant early star formation and late minor mergers are the most likely to have shallow inner dark matter density profiles in EDGE. However, none of the EDGE dwarfs considered here lie outside of the 68 percent confidence intervals of mass assembly histories in $\Lambda$CDM: they are all rather common. Rarer haloes could conceivably grow larger dark matter cores (though this is not guaranteed). We will consider this possibility in future work. \par

Alongside rarer assembly histories, we should consider also the sensitivity of dark matter core formation to the sub-grid baryonic physics model. \citet{2020arXiv200903313P} find that a very small increase in the burstiness of star formation (and, therefore, in the mass of inflowing and outflowing gas) can form a substantial dark matter core in an UFD -- rather similar to the Fcore model we have explored here -- without significantly increasing its stellar mass. Thus, it is possible that going to even higher numerical resolution, or including further physics not currently modelled in EDGE, could lead to the formation of larger dark matter cores in UFDs in $\Lambda$CDM.\par

\begin{figure}
\centering
\setlength\tabcolsep{2pt}%
\includegraphics[ trim={0cm 1cm 0cm 0cm}, clip=False, width=\columnwidth, keepaspectratio]{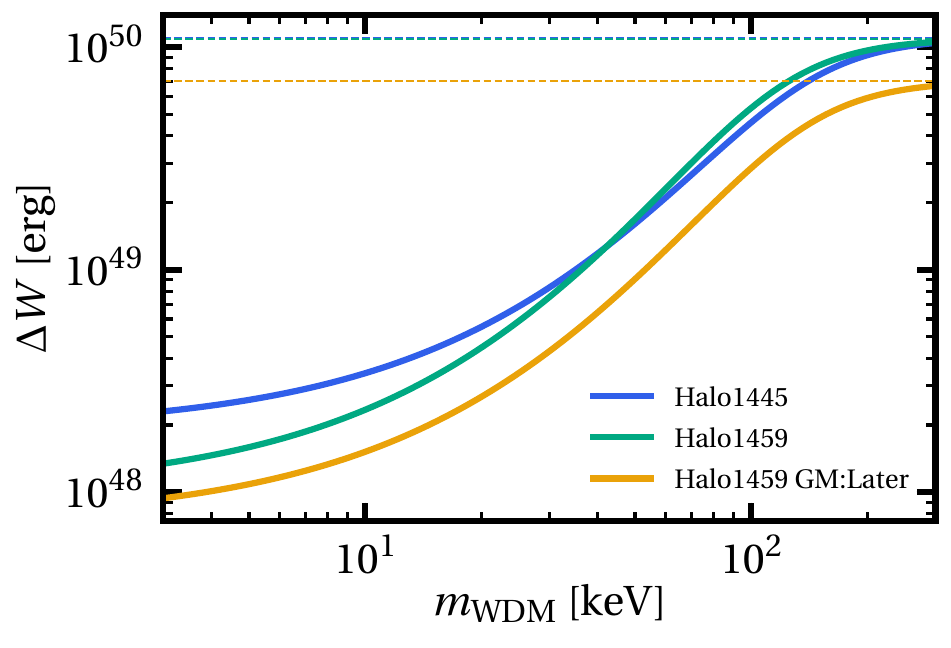}
\caption{The energy ($\Delta W$) required to transform a central cusp into coreNFW profiles fit to EDGE galaxies at $z=4$ (see Appendix \ref{app:energy} for details of how this is calculated). The halo concentration parameters have been varied, corresponding to WDM over a range of particle masses. Horizontal dashed lines mark the values of $\Delta W$ from Table \ref{tab:energetics}, which shows the energy required to lower the central density cusp in a CDM cosmology.} \label{fig:mWDM}
\end{figure}

\vspace{2mm}
\noindent
{\bf Full cores in $\Lambda$WDM $\mid$} A full dark matter core could also indicate a departure from a $\Lambda$CDM cosmology. One possibility is Warm Dark Matter (WDM). While WDM does not in itself produce significant dark matter cores, it does lower the concentration of dark matter haloes such that core formation via baryonic effects is more easily facilitated \citep{2012MNRAS.424.1105M, 2019MNRAS.490..962F}. In Figure \ref{fig:mWDM}, we quantify the energy required to form cores in $\Lambda$WDM as compared to $\Lambda$CDM, all other things being equal. Specifically, we show the energy difference ($\Delta W$, as calculated in Appendix \ref{app:energy}) between cuspy and dynamically heated potential profile fits, but varying the concentration parameter $c$ according to predictions for a thermal relic WDM particle \citep[as in][]{2012MNRAS.424..684S}. Notice that $\Delta W$ plummets by $\sim 2$ orders of magnitude for the lowest WDM particle masses (we truncate the figure at $m_{\rm WDM} = 3$\,keV since this is the conservative limit set by recent data constraints; e.g. \citealt{nadler21,enzi21,banik21}). However, there is an important caveat here that the reduction in central density in WDM as compared to CDM could also result in delayed star formation \citep[as in][]{2020MNRAS.498..702L} -- which would make core formation more difficult again. Secondly, the halo formation time, and therefore the final stellar mass, are different in a WDM cosmology \citep[i.e.][]{2014MNRAS.444.2333E, 2021arXiv210101198D}. We will consider these points using dedicated WDM simulations, in future work. \par

\vspace{2mm}
\noindent 
{\bf Full cores in $\Lambda$SIDM $\mid$} The presence of a dark matter core in EriII could point to a Self-Interacting Dark Matter (SIDM) cosmology. In some models, thermalisation occurs due to scattering between dark matter particles within dense halo centres. This leads to a density core that grows with time. \citet{2018MNRAS.481..860R} find that `coreNFW' density profiles \citep{2016MNRAS.459.2573R} provide an excellent match to the form of haloes in simulations of a velocity-independent SIDM cosmology, assuming interaction cross-sections within the range of current observational constraints. Using the methodology described in \S3.2 of \citet{2018MNRAS.481..860R}, which is calibrated to the  \citet{2012MNRAS.423.3740V} SIDM cosmological simulations, we find that an interaction cross-section of $>0.28\,\text{cm}^2/\text{g}$ is sufficient to lower the central density of our dwarf galaxy simulations as far as the Fcore model. Interestingly, this is consistent with the latest constraints on SIDM from dwarf galaxy scales \citep[e.g.][]{2018MNRAS.481..860R,correa21}.

In other SIDM models, core formation proceeds via dark particle annihilation in dense halo centres. In \citet{2021arXiv210100253Z}, an annihiliation cross-section is derived based on observational fits to the density profile of EriII assuming the velocity-independent SIDM model of \citet{2016JCAP...03..009L}. The lower-bounds of this fit are not dissimilar to the Fcore model. Following \citet{2021arXiv210100253Z}, the annihilation cross-section $\sigma/m = \Gamma/2v$, where $v$ is the typical velocity. We find that an annihilation cross-section of $\sigma/m < 1.7 \times 10^{-35} (f/10)^{-1} (v/10\,\text{km}^{-1})^{-1}\,\text{cm}^2\,\text{eV}^{-1}c^2$ is sufficient to lower the central density of our dwarf galaxy simulations as far as the Fcore model, which is roughly a factor of four greater than the 95 percent confidence constraint derived in \citet{2021arXiv210100253Z}. \par

\subsection{A recently accreted SC in EriII}\label{pathways2}

An alternative model to the ``full core'' scenario explored above is that EriII has a partial core. However, a SC in this model is unlikely to survive a Hubble time without falling into the potential centre. Instead, the SC \textit{must} be recently accreted, perhaps via a late minor merger. Then, an inflated half-light radius may be attained as the SC approaches tidal disruption. Such a donation may be entirely natural. Recall that galaxies drawn from our EDGE simulations were most similar to EriII’s observed properties if they assembled late from star-rich minor mergers (see \S\ref{sec:EDGE-Eri} and Table \ref{tab:eri_II_comparison}). In this scenario, the properties of EriII's SC may owe more to the initial conditions of its orbit and mass than to the density profile of EriII's inner dark matter halo.

Firstly, we note that even in this model a `pristine' dark matter cusp in EriII remains too destructive. As shown in Figure \ref{fig:dying_parameters}, the Halo1459 DMO model is unable to host SCs as large as that seen in EriII, even if only momentarily. A partial core, however, as in Halo1459GM:Later can support reasonably large, offset, SCs. As shown in Figure \ref{fig:dying_shape_analysis}, the tides experienced by SCs disrupting in this potential can naturally generate ellipticities similar to that seen in EriII's SC. However, this same model fails to produce a size or offset compatible with EriII’s SC (Figure \ref{fig:dying_parameters}). Furthermore, it yields tidal tails bright enough that we should have seen them in current data. Therefore, the notable absence of such tails appears to be a strong indicator that the SC in EriII is not consistent with this scenario. It may be worth considering whether tidal tails could be obscured. This might occur if the tails are viewed along the line-of-sight, if the luminosity of stripped stars is lower due to mass segregation within the SC, or if the tail material is spread over a wider track. The first scenario is possible, but demands fine-tuning and an unlikely chance orientation of the SC orbit. As shown in \citet{2018MNRAS.474.2479B}, mass segregation is greatest in denser SCs, which we find do not grow to the large size observed for the SC in EriII. The mass segregation for our more diffuse SCs has a negligible influence on tail brightness within the range of our luminosity cuts. \par

\begin{figure}
\centering
\setlength\tabcolsep{2pt}%
\includegraphics[ trim={0cm 1cm 0cm 0cm}, clip=False, width=\columnwidth, keepaspectratio]{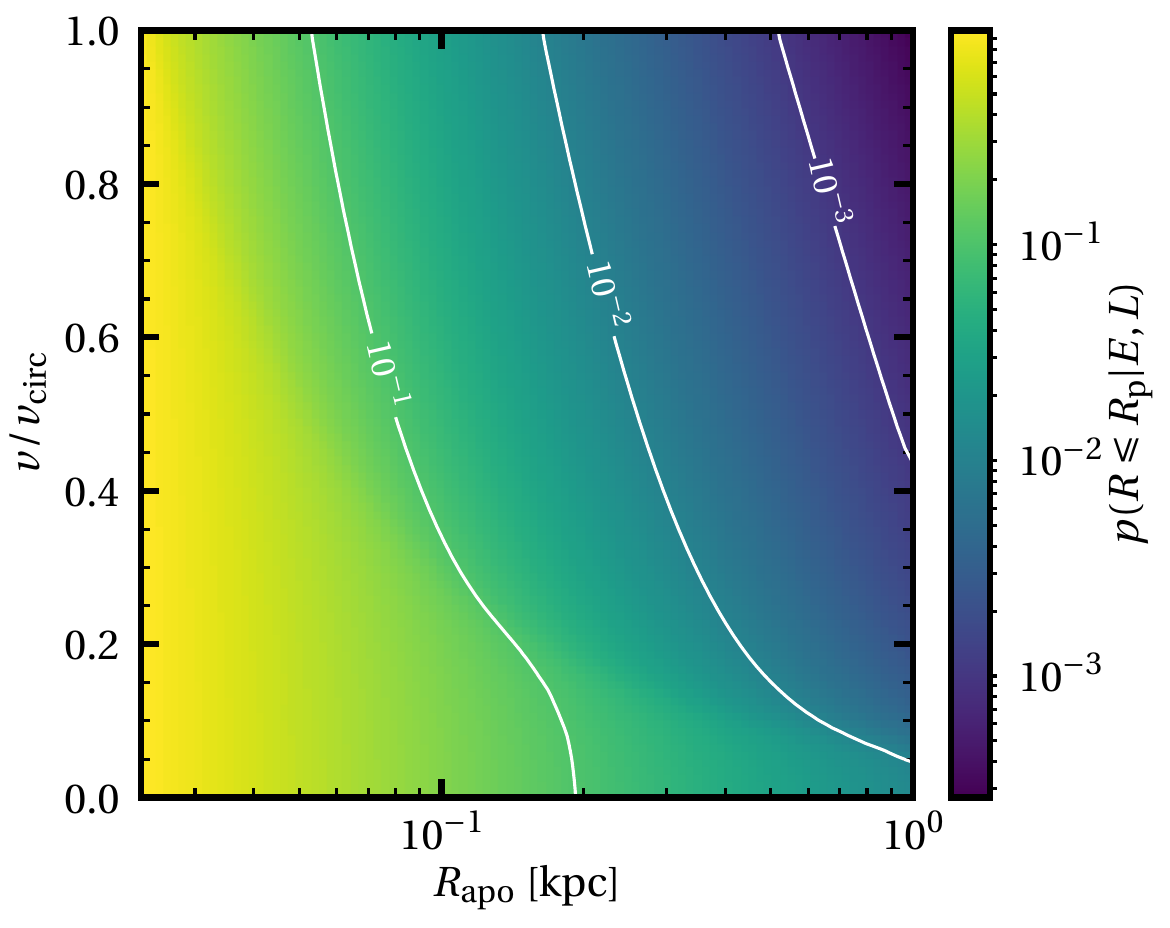}
\caption{The probability (Equation \ref{coleprob.eq}) of observing a SC at a projected radius $R\leqslant R_{\rm p}$, where $R_{\rm p}=23\,$pc, over a range of orbital velocities and apocentres. White lines show contours of constant probability.} \label{fig:cole_prob}
\end{figure}

We find that tidally-induced ellipticity occurs preferentially for higher apocentre orbits ($>100\,$pc) than the projected offset observed for EriII's SC. It is, therefore, reasonable to question the likelihood of observing EriII's SC at its current projected distance if the true separation from the centre of EriII is $>100\,$pc. To quantify this likelihood, we calculate the probability, described in \citet{2012MNRAS.426..601C}:
\noindent\begin{equation}
p(R\leqslant R_{\rm p}|E,L) = \int_{r_{\rm peri}}^{r_{\rm apo}}\left [ 1-(1-R^2_{\rm p}/r^2)^{1/2}_+ \right ] \frac{dr}{v_{r}} \bigg/ \int_{r_{\rm peri}}^{r_{\rm apo}}\frac{dr}{v_{r}},
\label{coleprob.eq}
\end{equation}
where $r_{\rm apo}$ and $r_{\rm peri}$ are the orbital apo- and pericentre, $v_{r}$ is the radial velocity and $(\cdot )_+ \equiv \max\{0,\cdot\}$. This gives the probability of observing an orbit at a projected radius $R\leqslant R_{\rm p}$, assuming a constant orbital energy $E$ and angular momentum $L$. In Figure \ref{fig:cole_prob}, we calculate $p(R\leqslant R_{\rm 23\,\text{pc}}|E,L)$ for a grid of orbits over a range of apocentres and initial tangential velocities. The background potential is fixed to our `Halo1459' Dehnen model, though the results are near-identical for our other background profiles. By construction, $p=1$ when $R\leqslant23$\,pc. However, $p$ rapidly drops to $\lesssim 0.1$ beyond $R \gtrsim 100\,$pc. \par

It may also be possible that a SC is so recently accreted to a ``partial-core'' host potential that it retains both a high primordial ellipticity and large half-light radius, and is not yet greatly disturbed by gravitational tides. We find that SCs with an initial half-light radius of $15\,$pc are extremely vulnerable to the tides within partial-core potentials, with only our most massive SCs ($M_{\rm ini} = 4.18\times10^{4}\,$M$_{\odot}$) surviving longer than a few Myrs. Such a SC could not be close to the centre of the host potential without experiencing significant tidal disruption, and as discussed above, the chance of observing a high-orbit realisation at the projected radius observed in EriII become increasingly unlikely at high radii. Determining a more exact likelihood would require simulations with higher initial orbital radii. \par

\subsection{Could EriII's SC be a luminous subhalo?}\label{pathways3}

In the end, neither the ``full core'' nor ``recent accretion'' scenarios are entirely satisfactory. Only the full core scenario is able to support in-situ SCs of the size seen in EriII. However, such a host potential may be challenging to explain in $\Lambda$CDM (see \S\ref{coreforming}). Furthermore, such a large and low density core is in mild tension with the observed stellar kinematics for EriII (see Figure \ref{fig:Dehnen_fit}). By contrast, the ``recent accretion'' scenario supports large SCs for our partial core and weakened cusp models, and naturally reproduces the ellipticity and orientation of the SC in EriII. But, it struggles to consistently support SCs as large as in EriII, and it predicts the formation of denser tidal tails that are absent in observations \citep{2020arXiv201200043S}. \par

However, it is interesting to note that the observed properties of the SC in EriII ($M_{v}=-3.6\pm0.6\,$mag; \citealp{2016ApJ...824L..14C}, $R_{\rm half}=23\pm3\,$pc; \citealp{2020arXiv201200043S}) place it in an uncertain regime between Milky Way Globular Clusters and potential dwarf galaxy candidates \citep[see e.g. Figure 2. of ][]{2019ARA&A..57..375S}. Therefore, perhaps the SC in EriII is not a true SC, but rather a luminous subhalo (or the nucleus of a luminous subhalo) -- the remnant of a dwarf galaxy merger. In this scenario, the subhalo's stellar distribution is shielded from the gravitational tides of EriII by its surrounding dark matter, similar to the behaviour of SCs in dark matter mini-haloes \citep{2021arXiv210403635B, 2021arXiv211201265V}. This could explain the lack of visible tidal tails, the comparatively large half-light radius and the high ellipticity, as the latter two are more natural properties of dwarf galaxies \citep{2008ApJ...684.1075M, 2017MNRAS.472.2670S, 2021MNRAS.503..176H}. Under this scenario, there are looser restrictions on both the infall time and slope of the host potential profile in EriII, so long as the dark matter envelope can effectively shield its stellar content. \par

With the simplistic assumption that haloes of mass $M_{200} <10^9\,\text{M}_{\odot}$ are well-fit by extrapolating stellar-mass halo-mass relations \citep[e.g.][]{2013MNRAS.428.3121M,read17}, a galaxy of $M_*=4.3\times10^3\text{M}_{\odot}$ would be hosted by a halo of approximately $M_{200} \sim 5.3\times10^8\text{M}_{\odot}$. Stars in a dispersion-supported dwarf galaxy inherit that velocity dispersion. Assuming that the velocity anisotropy has minimal impact within a small radius, the mass within the half-light radius can be estimated using the following formula \citep{2010MNRAS.406.1220W}:
\begin{equation}
    M_{1/2} = 3 G \left < \sigma^2_{\text{los}} \right > R_{1/2},
\end{equation}
where $\sigma_{\text{los}}$ is the line-of-sight velocity dispersion. This can be rearranged to give the velocity dispersion based on the half-light radius and corresponding enclosed mass. Assuming EriII’s SC (with half-light radius of $15\,$pc; \citealp{2020arXiv201200043S}) inhabits a dark matter subhalo with an NFW density profile, with a concentration at the median of the expected distribution in $\Lambda$CDM \citep{2017MNRAS.466.4974M}, this yields $\sigma_{\text{los}}\sim4.2\,\text{km}\,\text{s}^{-1}$. This is consistent with the constraints from the MUSE-Faint survey of $\sigma_{\text{los}} < 7.6\,\text{km}\,\text{s}^{-1}$ \citep{2020A&A...635A.107Z}. The dispersion is larger, however, than would be expected for a dark matter-free SC ($\sigma_{\text{los}} < 1.5\,\text{km}\,\text{s}^{-1}$; e.g. \citealt{2018MNRAS.476.3124C}) and so this scenario could be tested in future with more radial velocities for SC member stars.\par

Low-mass luminous subhaloes with properties comparable to the SC in EriII are present within the EDGE simulation suite (Taylor et al, in prep). However, there are no examples of such subhaloes which survive infall onto the EDGE galaxies presented here, perhaps due to a need to go to even higher resolution. We will explore such ideas in future work.\par

\section{Conclusions} \label{conclusions}

We have used a large suite of 960 direct $N$-body simulations of SCs orbiting within spherically symmetric profiles fit to UFD galaxies, simulated as part of the ‘Engineering Dwarfs at Galaxy Formation’s Edge’ (EDGE) project, to model the survival, evolution and properties of EriII's lone SC. We focussed, for the first time, on the puzzlingly large ellipticity ($0.31^{+0.05}_{-0.06}$; \citealt{2020arXiv201200043S}) of EriII's SC, asking the question: {\it can we form a SC as large, offset and elliptical as that in EriII within realistic UFDs in $\Lambda$CDM?} \par

As found in prior work, the large size and offset of EriII's SC are naturally explained if it orbits within a central dark matter core of size $\sim70\,\text{pc}$ and density $\lesssim2\times10^8\,\text{M}_{\odot}\,\text{kpc}^{-3}$. Such a flat dark matter core does not form naturally in the EDGE simulations and requires, therefore, either rarer assembly histories than explored in EDGE, additional physics currently missing from EDGE, or a shift to alternative dark matter models (see the discussion in \S\ref{coreforming}). Furthermore, due to the absence of tidal forces within the core, the cored model cannot explain the high ellipticity of EriII's SC through tidal stripping/deformation. However, this same lack of tidal forces causes any primordial birth ellipticity to become `frozen in' for long periods of time (see \S\ref{primordial_e} and Figure \ref{fig:primordial_e}). A high ellipticity may also be found if the SC orbit is observed close to edge-on (see Appendix \ref{app:edgeon}), though we find that this is unlikely to explain the ellipticity of the SC in EriII.  As such, in the ``full core'' model, the ellipticity of EriII's SC must reflect its birth ellipticity, which would be consistent with that of young SCs \citep{getman18}. \par

We also considered whether EriII's SC can be explained if it fell in recently and is currently on the verge of tidal disruption. This requires a central dark matter density for EriII that is not too-cusped and not-too-cored, exactly as predicted for late forming UFDs in the EDGE project (see discussion in \S\ref{pathways2}). A recent infall may be expected given that our EDGE simulations most similar to EriII assembled late from star-rich minor mergers. If the SC is observed during tidal disruption, then the orientation and ellipticity of EriII's SC are naturally reproduced. However, the primary problems with this model are that it struggles to explain the large size of EriII's SC and, in order to generate an ellipticity as high as that reported for EriII's SC, it produces more significant tidal tails that should have already been detected in current data (see \S\ref{pathways2}). Such a model requires fine-tuning of the orbit, history and current orientation of EriII’s SC.

A mixture of these scenarios may also provide possible solutions for the SC in EriII, but we stress that the presence of a core (or at least a partial-core) is a required feature in all models. Therefore, we currently favour the cored model. However, the ``recent accretion'' model could be salvaged if EriII's star cluster is actually a luminous dwarf galaxy and, therefore, tidally protected by a surrounding dark matter halo (see \S\ref{pathways3}). We will consider this idea in more detail in future work.\par

At present, the model that comes closest to explaining all the data for EriII's SC requires it to orbit in a large and low density dark matter core. Such a core is not expected in pure dark matter structure formation simulations in $\Lambda$CDM, though it could possibly be explained by baryonic effects (see the discussion in \S\ref{coreforming}). If baryonic effects cannot explain such a core, EriII's puzzling SC may call for us to move beyond the Cold Dark Matter model.\par

\section*{Acknowledgements}

MO acknowledges the UKRI Science and Technology Facilities Council (STFC) for support (grant ST/R505134/1).
This work was performed using the DiRAC Data Intensive service at Leicester, operated by the University of Leicester IT Services, which forms part of the STFC DiRAC HPC Facility (www.dirac.ac.uk). The equipment was funded by BEIS capital funding via STFC capital grants ST/K000373/1 and ST/R002363/1 and STFC DiRAC Operations grant ST/R001014/1. DiRAC is part of the National e-Infrastructure.
OA acknowledges support from the Knut and Alice Wallenberg Foundation and the Swedish Research Council (grant 2019-04659). 
This project has received funding from the European Union's Horizon 2020 research and innovation programme under grant agreement No.\ 818085 GMGalaxies. AP was supported by the Royal Society.
MR is supported by the Beecroft Fellowship funded by Adrian Beecroft and the Knut and Alice Wallenberg Foundation.
ET acknowledges the UKRI Science and Technology Facilities Council (STFC) for support (grant ST/V50712X/1).
MD acknowledges support by ERC-Syg 810218 WHOLE SUN.
We acknowledge Prof. Josh Simon for his helpful guidance, Prof. Mark Gieles for his useful advice on Globular Cluster evolution, Sebastiaan Zoutendijk for sharing his constraints on the density profile of Eridanus II, and the anonymous reviewer for their valuable comments.

In addition to software already mentioned, this work has made use of the public software {\sc Python} \citep{python}, {\sc Numpy} \citep{2011CSE....13b..22V}, {\sc Scipy} \citep{2020SciPy-NMeth} and {\sc Matplotlib} \citep{2007CSE.....9...90H}. \\

\section*{Author contributions}
The main roles of the authors were, using the CRediT (Contribution Roles Taxonomy) system\footnote{\url{https://authorservices.wiley.com/author-resources/Journal-Authors/open-access/credit.html}}: \par
MO: Conceptualisation; Data curation; Formal analysis; Investigation (lead); Writing – original draft. JR: Conceptualisation; Funding Acquisition; Project Administration; Resources; Writing – review and editing OA: Funding Acquisition; Methodology; Software; Writing – review and editing. AP: Conceptualisation; Funding
Acquisition; Methodology; Writing – review and editing. MR: Conceptualisation; Data curation; Methodology; Writing - review and editing. AG: Methodology. ET: Conceptualisation. SK:
Conceptualisation. MD: Software.

\section*{Data availability}
Data available upon reasonable request.

%%%%%%%%%%%%%%%%%%%%%%%%%%%%%%%%%%%%%%%%%%%%%%%%%%


\bibliographystyle{mnras}
\bibliography{sample} % if your bibtex file is called example.bib

%%%%%%%%%%%%%%%%%%%%%%%%%%%%%%%%%%%%%%%%%%%%%%%%%%

%%%%%%%%%%%%%%%%% APPENDICES %%%%%%%%%%%%%%%%%%%%%

%%%%%%%%%%%%%%%%%%%%%%%%%%%%%%%%%%%%%%%%%%%%%%%%%%

\appendix

\section{Shape fit with different methods} \label{appendix:a}

\begin{figure}
\centering
\setlength\tabcolsep{2pt}%
\includegraphics[ trim={0cm 1cm 0cm 0cm}, clip=False, width=\columnwidth, keepaspectratio]{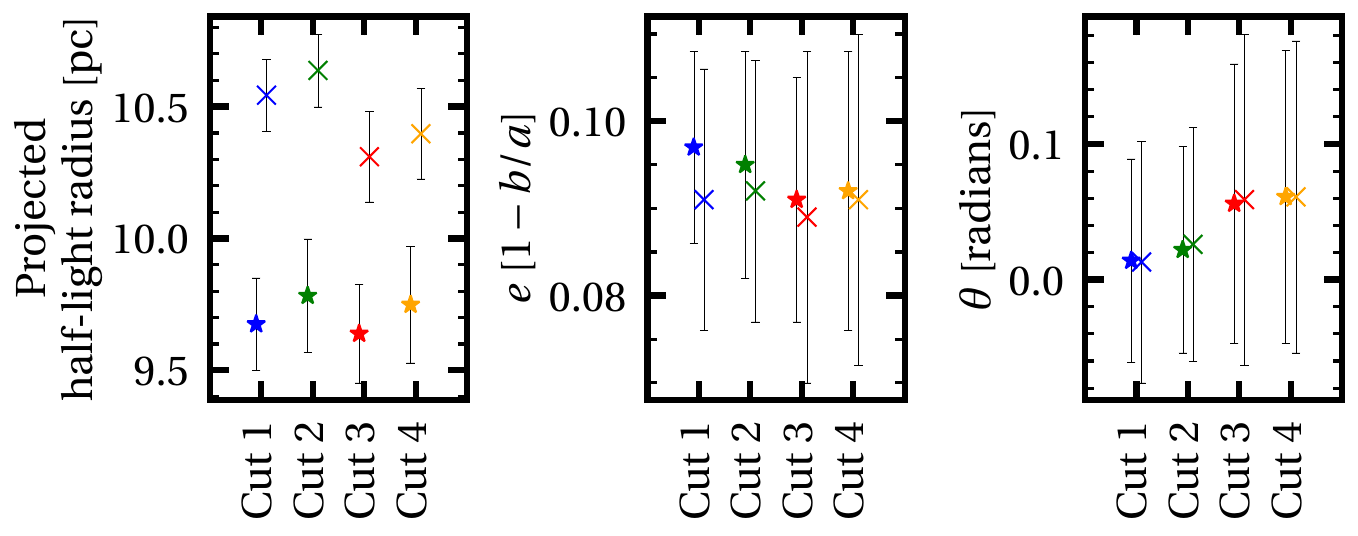}
\caption{Comparison of the projected half-light radius, ellipticity and orientation angle for a single example SC using different methods and luminosity cuts. Stars represent results using the direct analytical methods (direct calculation of the projected half-light radius and an iterative fit for the shape properties), whereas crosses represent results yielded by an MCMC method. The luminosity cuts are:
\textbf{Cut 1} (blue): $0.1 \geqslant L_{\rm V}/\text{L}_{\rm \odot,V} \geqslant 20$,
\textbf{Cut 2} (green): $0.1 \geqslant L_{\rm V}/\text{L}_{\rm \odot,V} \geqslant 100$,
\textbf{Cut 3} (red): $0.5 \geqslant L_{\rm V}/\text{L}_{\rm \odot,V} \geqslant 20$,
\textbf{Cut 4} (gold): $0.5 \geqslant L_{\rm V}/\text{L}_{\rm \odot,V} \geqslant 100$.
Error bars represent the one-sigma uncertainties.}
\label{fig:property_comparison}
\end{figure}

Throughout this paper, we estimate the shapes of SCs using the \texttt{analysis.halo.halo\_shape()} function in {\sc pynbody} (modified for 2D distributions), which iteratively solves the moment of inertia tensor. If the SC shape varies as a function of radius, as is the case for many of the simulations presented here, then the results depend on the size of the region that is fitted to. In our calculations, the definition of this region involves the half-light radius of the SC. This half-light calculation may itself be sensitive to the chosen luminosity cuts of the SC stars. \par

The size and shape of \textit{observed} stellar distributions are typically estimated using fits like in \citet{2008ApJ...684.1075M, 2016ApJ...833..167M}. To confirm that our numerical properties are analogous to such methods, we compare using a method developed by Goater et al (in prep). This method is based upon the affine-invariant Markov Chain Monte Carlo (MCMC) ensemble sampler {\sc emcee} \citep{2013PASP..125..306F}, and assumes a surface brightness profile which is fit to the simulated stellar data. Once appropriately normalised, it represents the mass-weighted probability of finding a star, $i$, at a given position, $r_i$, as,
\begin{equation}
    l_{i} = \frac{882 N}{{625\pi R_{1/2}^{2} (1-\epsilon)}} \exp\left(\frac{-42r_{i}}{25R_{1/2}}\right) \frac{m_{i}}{M},
\end{equation}
where $N$ is the number of stars in the sample, $M$ is the total stellar mass, $m_{i}$ the individual stellar mass, $\epsilon$ is the ellipticity defined as $\epsilon = 1 - b/a$, with $b/a$ as the minor-to-major-axis ratio of the system, $\theta$ is the position angle of the major axis, and $R_{\rm 1/2}$ is the projected half-light radius of the exponential radial profile. The elliptical radius is related to the stellar positions as:
\begin{equation}
    r_{i} = \left[\left(\frac{1}{1-\epsilon}(x_{i}\cos\theta - y_{i}\sin\theta)\right)^{2}+ \left(x_{i}\sin\theta + y_{i}\cos\theta\right)^{2}\right]^{\frac{1}{2}}.
\end{equation}
The total likelihood is the product of all probabilities, which is then calculated as:
\begin{equation}
    \ln \mathcal{L} = \sum_{i} \ln l_{i}.
\end{equation}
We place flat priors for $\epsilon$, $\theta$ and $R_{1/2}$ such that the $0 \leqslant \epsilon < 1$, $-\pi \leqslant \theta \leqslant \pi$, and $R_{1/2} > 0$. When fitting SCs we use the additional constraint that $R_{1/2} \leqslant 2\times R_{1/2\rm, numerical}$, where $R_{1/2\rm, numerical}$ is the numerically calculated projected half-light radius. Whilst this function is technically fit to the stellar mass distribution rather than the light distribution, it is often assumed that observed systems have uniform stellar populations unaffected by mass segregation \citep{2016ApJ...833..167M}. This approximation may be dubious for Globular Cluster systems \citep[see][]{1989ApJ...341..168B, 1998MNRAS.295..691B, 2009ApJ...700L..99A}. \par

In Figure \ref{fig:property_comparison}, we compare the size and shape for a single example SC as calculated using different methods and luminosity cuts. When using the MCMC method, we use $100$ walkers with $500$ burn-in steps and $2000$ main steps. The one-sigma uncertainties are similar between methods and luminosity cuts. The estimated ellipticity and orientation are almost identical between methods, however the MCMC method overestimates the projected half-light radius by a few sigma when compared to the direct analytical method. This overestimate can be very large for SCs that are close to dissolving, but is otherwise $\lesssim 1$\,pc across our simulation suite. Therefore, whilst the two methods are not formally in agreement, the expected systematic error of $\lesssim 1$\,pc does not influence our results. The choice of luminosity cut has a small effect on the size, ellipticity and orientation of the SC, but the magnitude of this effect is within the one-sigma uncertainties. \par

\section{Star cluster shape from an edge-on perspective}\label{app:edgeon}

\begin{figure}
\centering
\setlength\tabcolsep{2pt}%
\includegraphics[ trim={0cm 1cm 1cm 0cm}, clip=False, width=\columnwidth, keepaspectratio]{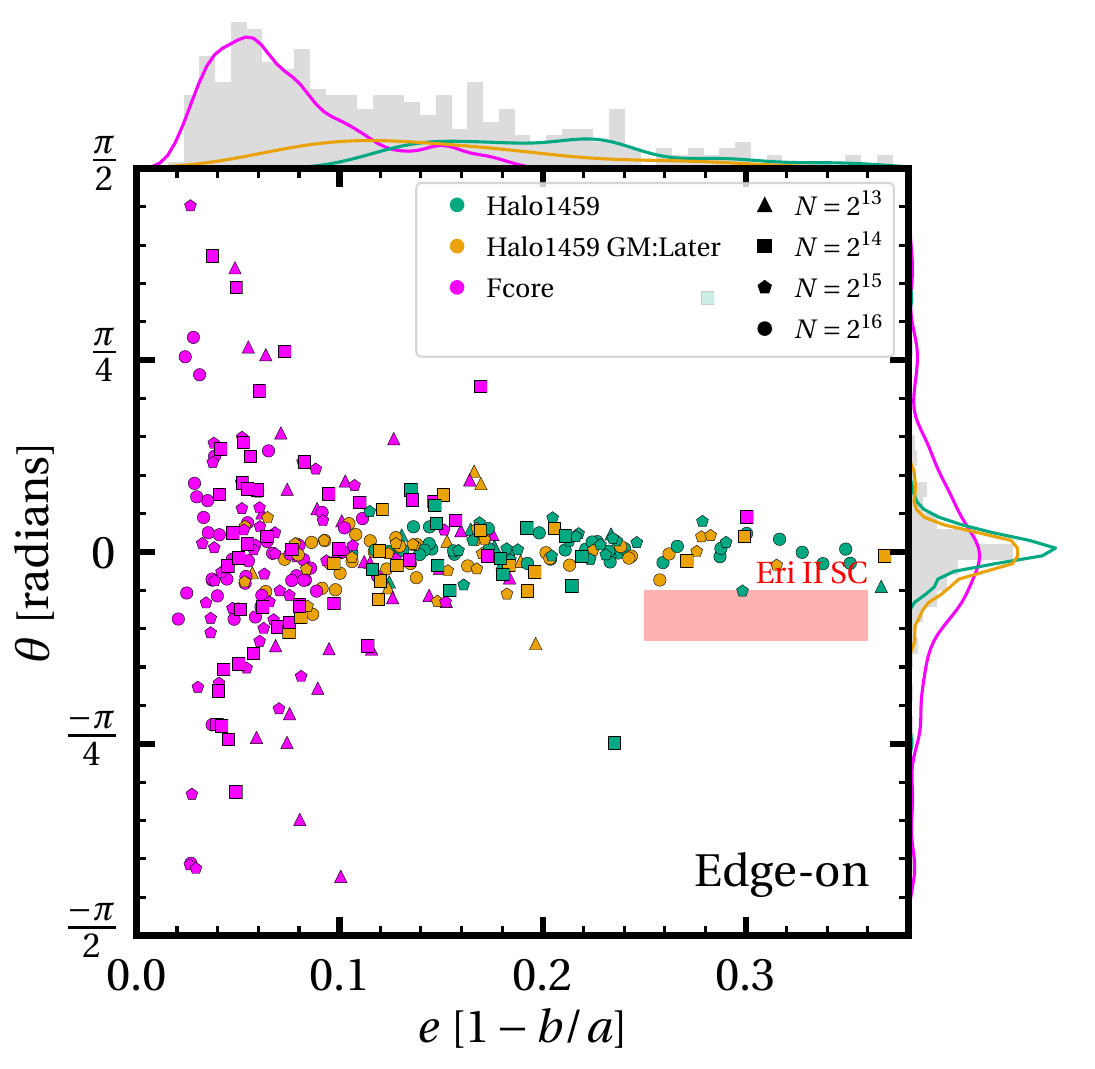}
\caption{The orientation $\theta$ versus ellipticity $e$ for all surviving SC simulations after a Hubble time, as in Figure \ref{fig:shape_analysis}, but as viewed edge-on to the orbital plane. A combination of tidal debris, which is denser when viewed edge-on to the orbit, and tidal compression leads to a greater ellipticity for many SCs.} \label{fig:shape_analysis_reorient}
\end{figure}

A SC in a cored potential can experience negligible gravitational tides in the face-on plane of its orbit, yet still experience a tidal compression across its $z$-axis \citep[see equation 21,][]{2011MNRAS.418..759R}. Therefore, it is necessary to consider the SC shape from both face-on and edge-on orientations for a complete understanding of the SC ellipticity. \par

In Figure \ref{fig:shape_analysis_reorient} we reconsider the SC shapes from Figure \ref{fig:shape_analysis}, but as viewed edge-on to the orbital plane. Now, there is a sharp spike of SCs at higher ellipticities. This is due in part to a tidal compression across the $z$-axis, but primarily to contamination from tidally stripped stars which form a dense track when viewed edge-on. If we exclude all unbound stars from our analysis, then only seven examples (in models Halo1459 and Halo1459 GM:Later) exceed $e=0.15$. The ellipticities of SCs in the Fcore potential remain low in almost all cases, because the flatter potential liberates fewer stars. A single data point in the Fcore potential achieves an ellipticity comparable to that of the SC in EriII, but it is a near-destroyed SC that does not match EriII in any other respect. \par

These increased ellipticities are predominantly a consequence of our idealised simulations, where the SC orbit remains in the same $x-y$ plane and the unbound stars remain dynamically cold for a full Hubble time. Even then, the SC must be viewed extremely close to edge-on before the perceived ellipticity increases notably. \par

\section{Energy requirements for dark matter core formation in the EDGE UFDs}\label{app:energy}

\begin{table}
\centering
\begin{tabular}{lccc} 
\toprule
\textbf{Name} & \textbf{$E_{\rm SN}$ [erg]} & \textbf{$\Delta W$ [erg]} & \textbf{$\epsilon_{\rm DM}$} \\
\midrule
\textcolor{Halo1}{Halo1445} & $7.28 \times 10^{53}$ & $1.10 \times 10^{50}$ & $1.51 \times 10^{-4}$ \\
\textcolor{Halo2}{Halo1459} & $1.17 \times 10^{54}$ & $1.09 \times 10^{50}$ & $9.37 \times 10^{-5}$ \\
\textcolor{GM}{Halo1459 GM:Later} & $2.72 \times 10^{53}$ & $7.07 \times 10^{49}$ & $2.60 \times 10^{-4}$ \\
\bottomrule
\end{tabular}
\caption{Comparison of the estimated total SNe energy ($E_{\rm SN}$), the energy required to lower the central density cusp (cusp gravitational binding energy minus core gravitational binding energy, $\Delta W$), and the coupling efficiency ($\epsilon_{\rm DM}$) for the chosen EDGE haloes at $z=4$. For the calculation of $E_{\rm SN}$ we consider only stars within the core radius (taken from coreNFW profile fits, \citealp{2016MNRAS.459.2573R}), as stars exterior to this radius do not contribute to the heating of the dark matter cusp.}
\label{tab:energetics}
\end{table}

We perform a simple comparison of the energy required to unbind our dark matter cusps with the available energy provided by SNe explosions, following prior work in \citet{2012ApJ...759L..42P, 2016MNRAS.459.2573R, 2018MNRAS.476.3124C, 2020arXiv201005930T}. We make this comparison at $z=4$ because star formation has permanently quenched by this time, noting, however, that this does not account for the additional late-time coring due to minor mergers in Halo1459 GM:Later. \par

We estimate the SNe energy from the IMF as follows:
\noindent\begin{equation}
E_{\rm SF} = \frac{M_{*} \xi E_{\rm SN}}{\left \langle m_{*} \right \rangle},
\label{SNe_energy.eq}
\end{equation}
where $M_{*}$ is the total stellar mass of the galaxy at $z=4$, $\left \langle m_{*} \right \rangle$ is the expectation value for the stellar mass, $\xi$ is the number fraction of stars that become SNe ($m_{*} > 8\,\mathrm{M}_{\odot}$) and $E_{\rm SN} = 10^{51}\,$erg is the energy per SNe. Our implementation of {\sc ramses} utilizes a Chabrier IMF \citep{2003PASP..115..763C}, for which we evaluate $\left \langle m_{*} \right \rangle$ and $\xi$ by integrating over the range $0.1 < m_{*}/\mathrm{M}_{\odot} < 120$ as 0.680 and 0.00802 respectively. This is an imperfect representation of the SNe energy because it does not consider that some SNe could have occurred whilst an accreted stellar particle was still ex-situ, and may not have contributed directly to the dynamical heating of the main progenitor. \par

The energy needed to depress the central density can be found by analysing the enclosed mass profiles of the haloes at $z=4$. We fit a coreNFW profile \citep{2016MNRAS.459.2573R} where the concentration parameter $c$ is constrained based on fits from \citet{2014MNRAS.441.3359D}. A cuspy profile is approximated by setting the parameter $n=0$ (equivalent to an NFW profile \citealp{navarro1997universal}). Then, the energy difference $\Delta W$ between the cuspy NFW and fitted coreNFW profiles is calculated as in Equation 22 of \citet{2016MNRAS.459.2573R}:
\noindent\begin{equation}
\Delta W = -\frac{1}{2} \int_{0}^{\infty} \frac{G(M^2_{\rm NFW} - M^2_{\rm coreNFW})}{r^2} dr.
\label{deltaW.eq}
\end{equation}
A coupling efficiency between the SNe energy and dark matter is defined as $\Delta W / E_{\rm SF}$. The results are summarised in Table \ref{tab:energetics}. \par

Although Halo1459 and Halo1459 GM:Later are based upon the same modified initial conditions, Halo1459 GM:Later requires less energy to unbind its dark matter cusp. This is because the later assembly of Halo1459 GM:Later has left it with a diminished central density at the earliest times. Halo1445 exhibits a similar coupling efficiency to Halo1459 GM:Later, which also owes to its delayed mass growth. \par

% Don't change these lines
\bsp	% typesetting comment
\label{lastpage}
\end{document}